\documentclass[prl,twocolumn,aps]{revtex4-1}

\usepackage{color}
\usepackage{graphicx}
\usepackage{subfigure}
\usepackage{afterpage}
\usepackage{dcolumn}
\usepackage{bm}
\usepackage{color}
\usepackage{float}
\usepackage{amsmath}
\usepackage{bm}
\usepackage{amssymb}
\usepackage{siunitx}
\usepackage{cleveref}
\usepackage{natbib}

\def\url#1{\expandafter\string\csname #1\endcsname}

\newcounter{defcounter}
\usepackage{amsthm}

\newenvironment{myequation}{
\addtocounter{equation}{-1}
\refstepcounter{defcounter}

\begin{equation}}
{\end{equation}}

\begin{document}

\title{Edge Solitons in a Nonlinear Mechanical Topological Insulator}

\author{David D. J. M. Snee$^{\small{1},\ast}$}
\author{Yi-Ping Ma$^{\small{1},\dagger}$}
\affiliation{$^{\small{1}}$Department of Mathematics, Physics and Electrical Engineering, Northumbria University, Newcastle Upon Tyne, NE1 8ST, UK
}

\begin{abstract}
We report localized and unidirectional nonlinear traveling edge waves discovered theoretically and numerically in a 2D mechanical (phononic) topological insulator. The lattice consists of a collection of pendula with weak Duffing nonlinearity connected by linear springs. We show that the classical 1D nonlinear Schr\"odinger equation governs the envelope of 2D edge modes, and study the propagation of traveling waves and rogue waves in 1D as edge solitons in 2D. As a result of topological protection, these edge solitons persist over long time intervals and through irregular boundaries.
\end{abstract}

\maketitle

Topological insulators (TIs) are a unique state of matter with behavior that has significant ramifications in the fields of both condensed matter physics and optics. The importance of TIs has been established in the condensed matter community for almost a decade, with extensive literature on both one-dimensional and multi-dimensional systems \cite{Moore-2010,Hasan-Kane-2010,Qi-Zhang-2010}. The one stand out property of TIs which holds the ongoing interest in the topic is that a clear dichotomy exists between the edge (surface) and the bulk of the material as electrons are conducted only on the edge whilst the bulk is insulating. The existence of such edge states at the interface between two bulk materials with different topological invariants is guaranteed by the principle of bulk-edge correspondence \cite{Qi-Wu-Zhang-2006-PRB,CTSR16-RMP}. More recently, the theoretical framework underlying quantum TIs has been generalized to photonic systems governed by classical electromagnetic fields \cite{Haldane-Raghu-PRL-2008,Ling-etal-2014,Ozawa-etal-2018}. In analogy to electrons in traditional TIs, electromagnetic waves in photonic TIs propagate along the edge with very little backscattering, even in the presence of disorders such as missing site(s) on the edge of a photonic lattice \cite{Wang-etal-2009,Khanikaev-etal-2013,Rechtsman-etal-2013}.

The emerging field of topological mechanics utilizes such topological principles to reveal new collective excitations in classical mechanical (phononic) systems \cite{Huber-2016-NP,MaSh16_SciAdv}. These topological acoustic metamaterials can be classified into two families depending on whether the topological edge modes appear at zero frequency or high frequencies. In the zero frequency case, these edge modes are identified as floppy modes and self-stress states in Maxwell frames \cite{Kane-Lubensky-2014}. Here we focus on the high frequency case, where topologically protected transport via phonons is enabled. Seminal work in this direction includes analogues of the quantum Hall effect using a lattice of hanging gyroscopes \cite{NKRVTI_PNAS_2015}, and the quantum spin Hall effect (QSHE) using a lattice of coupled pendula \cite{Susstrunk-Huber-2015} and bi-layered lattices of disks and springs \cite{Pal-etal-2016}.

\begin{figure}[b]
    \hspace*{-1cm}\includegraphics[width=.35\textwidth]{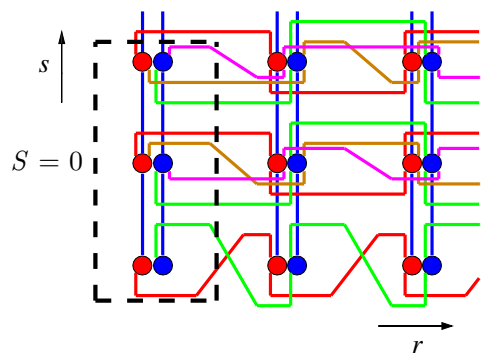}
    \caption{(Color online) Schematic view of the first unit cell $S=0$ consisting of three sites on a generic lattice. Each site hosts an $x$ (red) and a $y$ (blue) pendulum carrying a mass. Simple springs in the $s$-direction connecting pendula are shown as blue lines. Couplings between pendula in the $r$-direction are shown as red and green lines. Cross couplings between the $x$ and $y$ pendula of neighboring sites are shown as brown and magenta lines. Negative couplings are realized via one lever arm, and positive couplings via two lever arms.}
    \label{fig:lattice-layout}
\end{figure}

Over the last century, the theory of nonlinear waves in continuous systems has formed a cornerstone of nonlinear science \cite{Ablowitz-2013}. A fundamental result is that in such systems, dispersion can balance with nonlinearity to produce robust localized nonlinear traveling waves known as solitons. The theory of nonlinear waves in discrete systems has flourished more recently, especially in the context of nonlinear optics and Bose-Einstein condensates \cite{Kartashov-etal-2011}. In mechanical lattices, the study of nonlinear waves dates back to the classical 1D Fermi-Pasta-Ulam-Tsingou (FPUT) lattice \cite{Fermi-etal-1974}, with its 2D extension still being analyzed quite recently \cite{Butt-Wattis-2007,Chen-Hermann-2017}. Coupled pendula have been extensively studied in terms of synchronization and destabilization \cite{PiRo03_CUP}, while coherent structures therein such as breathers continue to capture considerable interest \cite{CuEn09_PRL,XuAl14_PRE,PaHa16_PLA}. Nonlinear elastic metamaterials known as granular crystals have attracted much recent attention \cite{ChPo17_JoPCM}, with topological effects already demonstrated \cite{CKTKY17_PRL}. The existing literature on nonlinear topological mechanics is quite limited, but notably includes establishment of topological solitons as the nonlinear mechanism for zero-frequency floppy modes to propagate through the bulk of a 1D Maxwell frame \cite{Chen-etal-2014}, and the effect of nonlinearity on the resonant characteristics of high-frequency edge modes in both 1D and 2D mechanical TIs \cite{Pal-etal-2018}.

In this Letter, we use the theory of nonlinear waves to describe nonlinear interactions between high-frequency edge modes in a 2D mechanical TI. The primary aim is to discover topologically protected edge solitons (TPES), which are nonlinear traveling edge waves that inherit the topological protection of the corresponding linear system. Using dimension reduction and asymptotic analysis, the original 2D nonlinear system is reduced to the classical 1D nonlinear Schr\"odinger (NLS) equation \cite{Ablowitz-Prinari-2008}. This equation admits a plethora of soliton solutions, all of which correspond to TPES in the original system. Theoretical predictions are compared to numerical simulations on relatively large domains with excellent agreement. This practically enables robust transport of a mechanical state from one location to another on the edge of a generic 2D lattice. We highlight that related methods have led to the discovery of TPES in photonic Floquet TIs \cite{Ablowitz-Curtis-Ma-2014,Ablowitz-Ma-2015,Leykam-Chong-2016}, polariton TIs \cite{KaSk16_Optica,GuYu17_SciRep,LiYe18_PRB}, and classical optical networks \cite{ShKi17_PNAS}, so our work may be viewed as a natural sequel into the mechanical (phononic) world.

The 2D mechanical TI we study is the one proposed by S{\"u}sstrunk and Huber \cite{Susstrunk-Huber-2015}, which is the first mechanical implementation of the QSHE. In the QSHE, two counter-propagating helical edge modes traverse the edge similarly with the exception of their spin degree of freedom \cite{Kane-Mele-2005,Bernevig-Zhang-2006}. In \cite{Susstrunk-Huber-2015}, a key analogy is formed between the quantum mechanical description of a TI and its classical mechanical equivalent, which in turn allows for a comparison of the two at the dynamical level: in both cases the eigenstates of the matrix encoding the lattice problem describes the QSHE. Following \cite{Susstrunk-Huber-2015}, we consider a 2D lattice indexed by $(r,s)$, where each site contains two pendula $x_{r,s}$ and $y_{r,s}$ that swing only in the $s$-direction. Linear springs are attached to the pendula to achieve couplings in both $s$- and $r$-directions. In \cite{Susstrunk-Huber-2015}, linear equations of motion based on Newton's second law are presented for this system; here we include an additional cubic (Duffing) nonlinearity inherent to pendula.

For this particular system, it is most convenient to group the sites in the $s$-direction into unit cells indexed by $S$, with each cell consisting of three sites $(x_{r,S}^{(j)},y_{r,S}^{(j)})$, $j=0,1,2$. Figure~\ref{fig:lattice-layout} shows a schematic view of the couplings on a generic lattice with the first unit cell ($S=0$) emphasized, where the first row represents $(x_{r,S}^{(0)},y_{r,S}^{(0)})$, the second row represents $(x_{r,S}^{(1)},y_{r,S}^{(1)})$, and the third row represents $(x_{r,S}^{(2)},y_{r,S}^{(2)})$. The detailed equations of motion are presented in Supplementary Material \cite{Supplementary-Material}; hereafter we consider the compact matrix form
\begin{equation}\label{eq:eqn-of-motion}
\boldsymbol{\ddot{X}}_{r,S}(t) = (\boldsymbol{\mathcal{L}}\boldsymbol{X})_{r,S} + \sigma \boldsymbol{\mathcal{N}}_{r,S}
\end{equation}
where $\boldsymbol{X}=[x^{(0)}, y^{(0)}, x^{(1)}, y^{(1)}, x^{(2)}, y^{(2)}]^{T}$, $t$ denotes time, $\cdot$ denotes time derivative, $\boldsymbol{\mathcal{L}}$ is the linear operator encoding the couplings of the system, $\sigma$ is the strength of the Duffing nonlinearity, and $\boldsymbol{\mathcal{N}}=\boldsymbol{X}^3$. Unless otherwise specified, the 2D domain is taken to be rectangular with $N_r$ sites in the $r$-direction, $r=0,1,2,\cdots,N_r-1$, and $N_S$ unit cells in the $S$-direction, $S=0,1,2,\cdots,N_S-1$. On a 1D domain with $\mathcal{L}$ chosen as the discrete Laplacian, Eq.~\eqref{eq:eqn-of-motion} reduces to the 1D nonlinear Klein-Gordon equation with a cubic nonlinearity \cite{Ginibre-Velo-1985}, from which the 1D NLS equation can be derived using the method of multiple scales \cite{Ablowitz-2013}. Here we combine this procedure with dimension reduction to derive a 1D amplitude equation from the general 2D system given by Eq.~(\ref{eq:eqn-of-motion}).

In the linear problem ($\sigma=0$), to describe an edge state of a specific boundary, say along $S$, we take the Fourier transform $\boldsymbol{X}_{r,S}(t)=e^{i\theta}\boldsymbol{X}_r^{E}+c.c$, $c.c$ denoting complex conjugacy, where the exponent $\theta=Sk-t\alpha(k)$, with $k$ being the wavenumber in $S$ (along the edge), $\alpha(k)$ being the dispersion relation, and $\boldsymbol{X}_r^{E}$ being the 1D edge state that decays in $r$ (perpendicular to the edge). This reduces Eq.~(\ref{eq:eqn-of-motion}) to the eigenvalue problem
\begin{equation}\label{eq:linear-eigenvalue-problem}
\boldsymbol{\mathcal{L}}(k)\boldsymbol{X}_r^{E} =-\alpha(k)^2\boldsymbol{X}_r^{E},
\end{equation}
where $\boldsymbol{\mathcal{L}}(k)$ denotes the 1D linear operator after the Fourier transform, and $\boldsymbol{X}_r^{E}$ is normalized such that $\|\boldsymbol{X}_r^{E}\|^2_2=\sum_j |\boldsymbol{X}_{j}^{E}|^2=1$. In the nonlinear problem ($\sigma\neq0$), similarly to \cite{Ablowitz-Ma-2015,Ablowitz-Curtis-Ma-2014}, we construct a weakly nonlinear edge mode by introducing a spectral envelope with a narrow width $0<\epsilon\ll 1$ at carrier wavenumber $k=k_0$ and carrier frequency $\alpha_0\equiv\alpha(k_0)$, in the reference frame co-moving with the group velocity of the envelope $\alpha_0'\equiv\alpha'(k_0)$. Specifically, we choose a multiple scale ansatz of the form
\begin{equation}\label{eq:ansatz}
\boldsymbol{X}(r,S,\tilde{S},t,\tau)=\epsilon\left\{C(\tilde{S},\tau)e^{i(Sk_0-t\alpha_0)}\boldsymbol{X}_r^{E}+c.c\right\}+h.o.t,
\end{equation}
where $C(\tilde{S},\tau)$ is the scalar envelope function and evolves in the slow variables $\tilde{S}\equiv\epsilon( S - \alpha_0't)$ and $\tau\equiv\epsilon^2 t$, and $h.o.t$ denotes higher order terms in $\epsilon$. Thus, the time and space differential operators are now written as $\partial_t=\epsilon^2\partial_{\tau}-\epsilon\alpha'_0\partial_{\tilde{S}}-\alpha_0i$ and $\partial_S=\epsilon\partial_{\tilde{S}}+k_0i$ respectively.

\begin{figure}[t]
\centering
    \includegraphics[width=8.7cm,height=4cm]{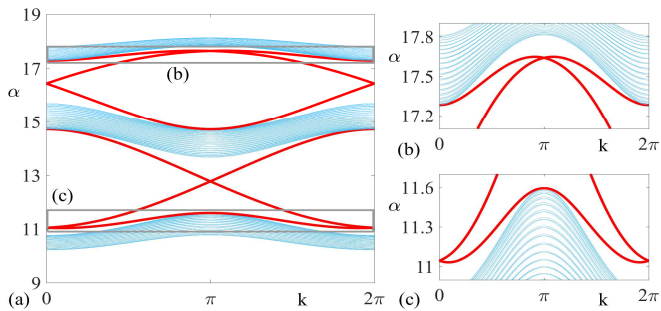}
    \caption{(Color online) The dispersion relation $\alpha(k)$ computed on the left/right edge of a rectangular lattice with $N_r=60$ sites in the $r$-direction. The bulk spectrum is shown in blue and the edge spectrum is shown in red.}
    \label{fig:dispersion-relation}
\end{figure}

The dispersion relation $\alpha(k)$, $k \in [0, 2\pi)$, numerically computed by solving Eq.~(\ref{eq:linear-eigenvalue-problem}) on a finite 1D domain, is shown in Fig.~\ref{fig:dispersion-relation}. The bulk (continuous) spectrum consists of three bands (blue), and the edge (point) spectrum exist in the two gaps (red). Thus, for each wavenumber $k$ there are precisely six point eigenvalues, three in either gap, whose corresponding eigenfunctions are the edge states that decay in $r$. The topologically protected edge states are precisely those in the bulk band gap, namely the range of $\alpha$ where no bulk state exists.

One can now substitute the ansatz \eqref{eq:ansatz} directly into Eq.~\eqref{eq:eqn-of-motion} and expand asymptotically in $\epsilon$. In particular, we expand the dispersion relation $\alpha(k)$ around the carrier wavenumber $k_0$ in operators form as
\begin{equation}\label{eq:dispersion-taylor-expansion}
\alpha(k)=\alpha_0-i\alpha'_0\epsilon\partial_{\tilde{S}}-\frac{1}{2}\alpha''_0\epsilon^2\partial_{\tilde{S}\tilde{S}}+O(\epsilon^3),
\end{equation}
where $\alpha_0''\equiv\alpha''(k_0)$. At $O(\epsilon)$ and $O(\epsilon^2)$ the resulting terms vanish producing trivial solutions. To leading nontrivial order, at $O(\epsilon^3)$, taking the inner product of both sides of the equation with $\boldsymbol{X}_r^{E}$, where the inner product is defined as $\langle\boldsymbol{g},\boldsymbol{h} \rangle=\sum_j g_j^*h_j$, then the 1D classical second-order NLS equation appears in canonical form:
\begin{equation}\label{eq:NLS}
iC_{\tau} + \frac{\alpha''_0}{2}C_{\tilde{S}\tilde{S}}+\frac{3\tilde{\sigma}}{2\alpha_0}|C|^2C=0,
\end{equation}
where $\tilde{\sigma}=\sigma \|\boldsymbol{X}_r^{E}\|^4_4\equiv\sigma\sum_j |\boldsymbol{X}_{j}^{E}|^4$. The NLS equation is a maximally balanced equation and is focusing or defocusing if $\alpha''_0\tilde{\sigma}/\alpha_0 > 0$ or $\alpha''_0\tilde{\sigma}/\alpha_0 < 0$ respectively.

The propagation of different types of edge solitons around the boundaries of the mechanical lattice can now be investigated by numerically solving Eq.~\eqref{eq:eqn-of-motion} with the following initial condition resulting from the ansatz (\ref{eq:ansatz})
\begin{equation}\label{eq:general-initial-condition}
\boldsymbol{X}_{r,S}(t=0)=\epsilon C(S) e^{iSk_0}\boldsymbol{X}_r^{E}+c.c,
\end{equation}
where $C(S)$ represents the initial envelope needed to produce the desired soliton solution to Eq.~(\ref{eq:NLS}). Hereafter we take $\epsilon=0.1$ and explore differing values of $k_0$.

\begin{figure}[h!]
  \centering
  \subfigure[]{\includegraphics[width=.2\textwidth]{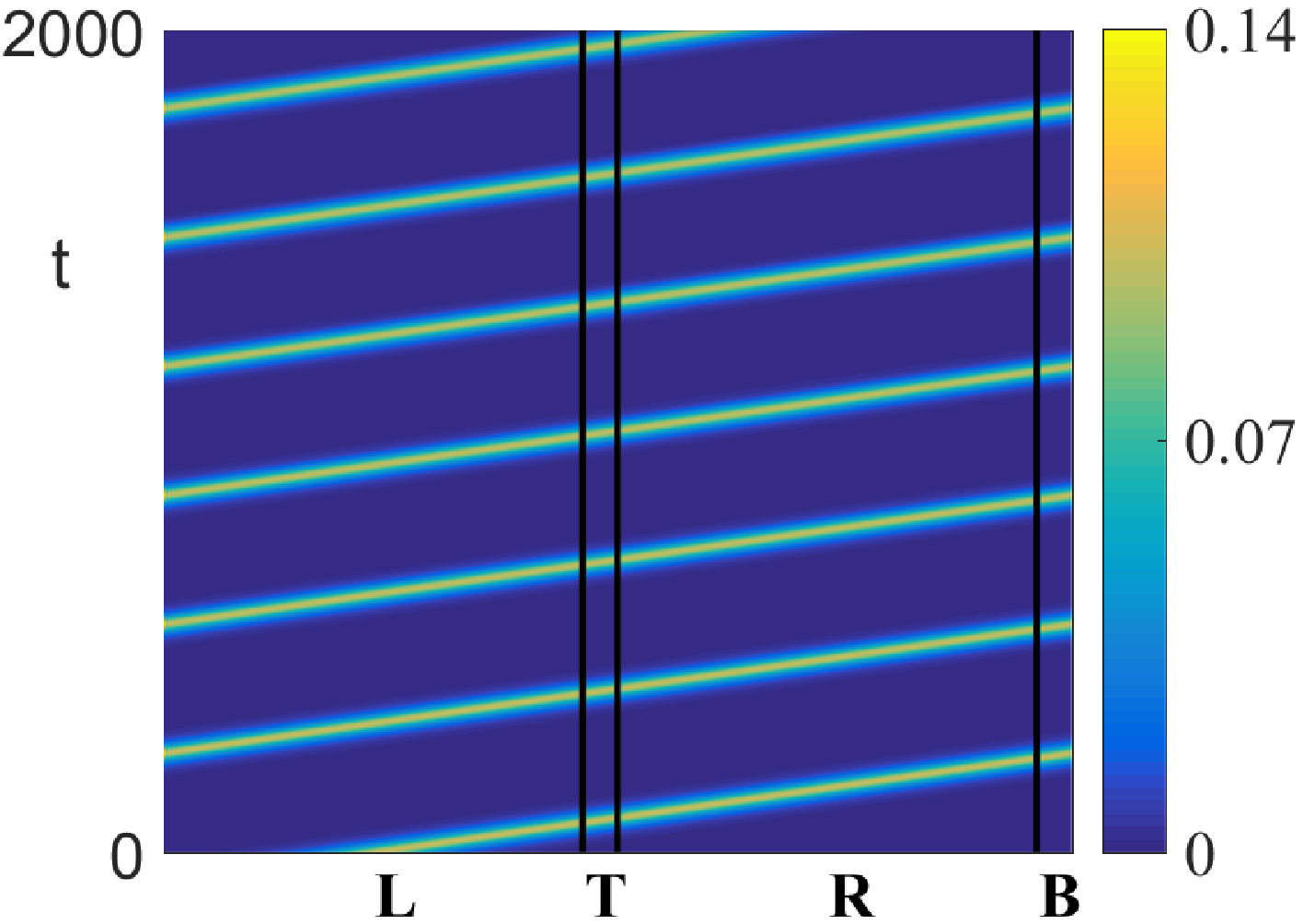}\label{fig:space-time-bright}}\hfill
  \subfigure[]{\includegraphics[width=.2\textwidth]{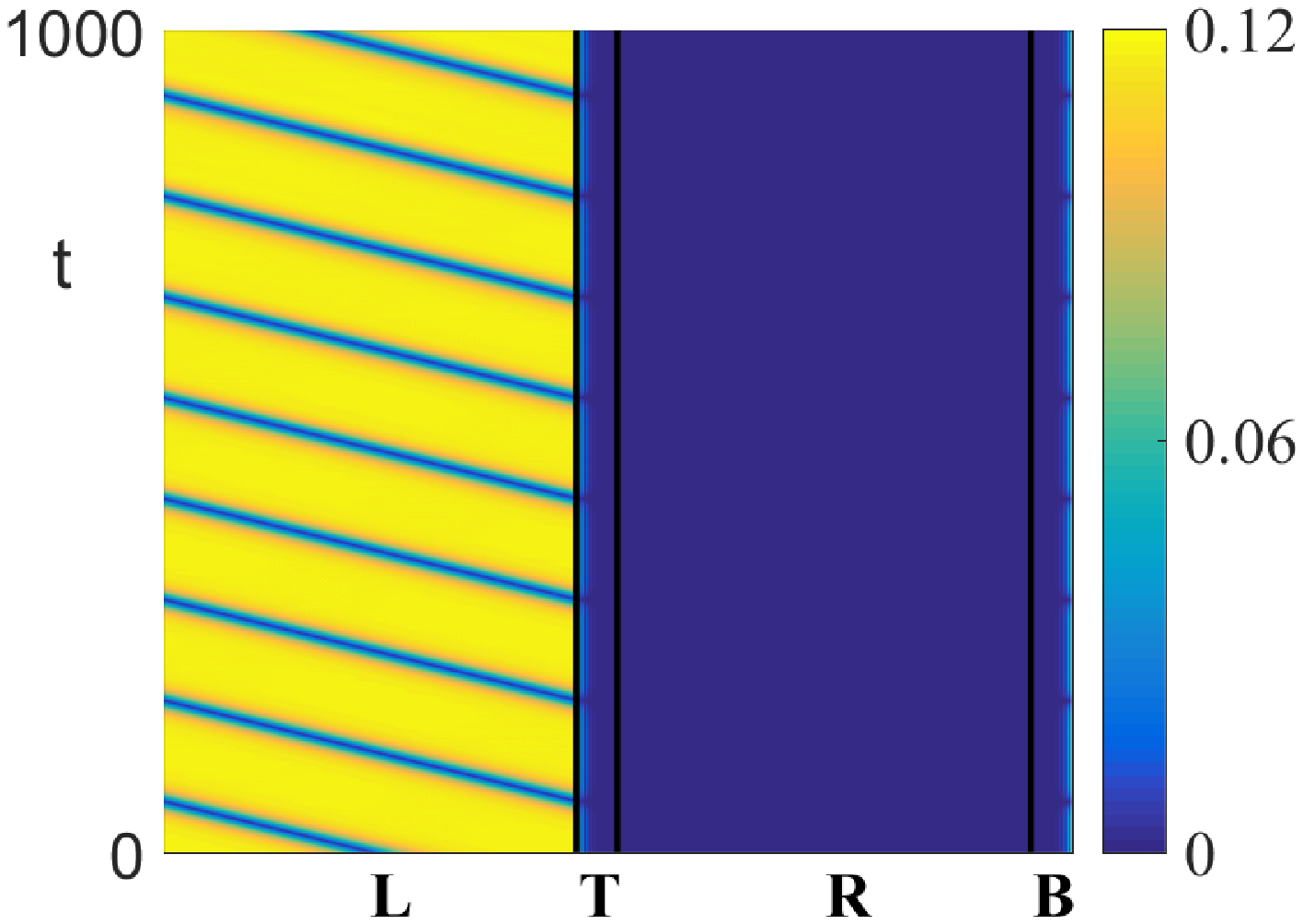}\label{fig:space-time-dark}}
    \subfigure[]{\includegraphics[width=.2\textwidth]{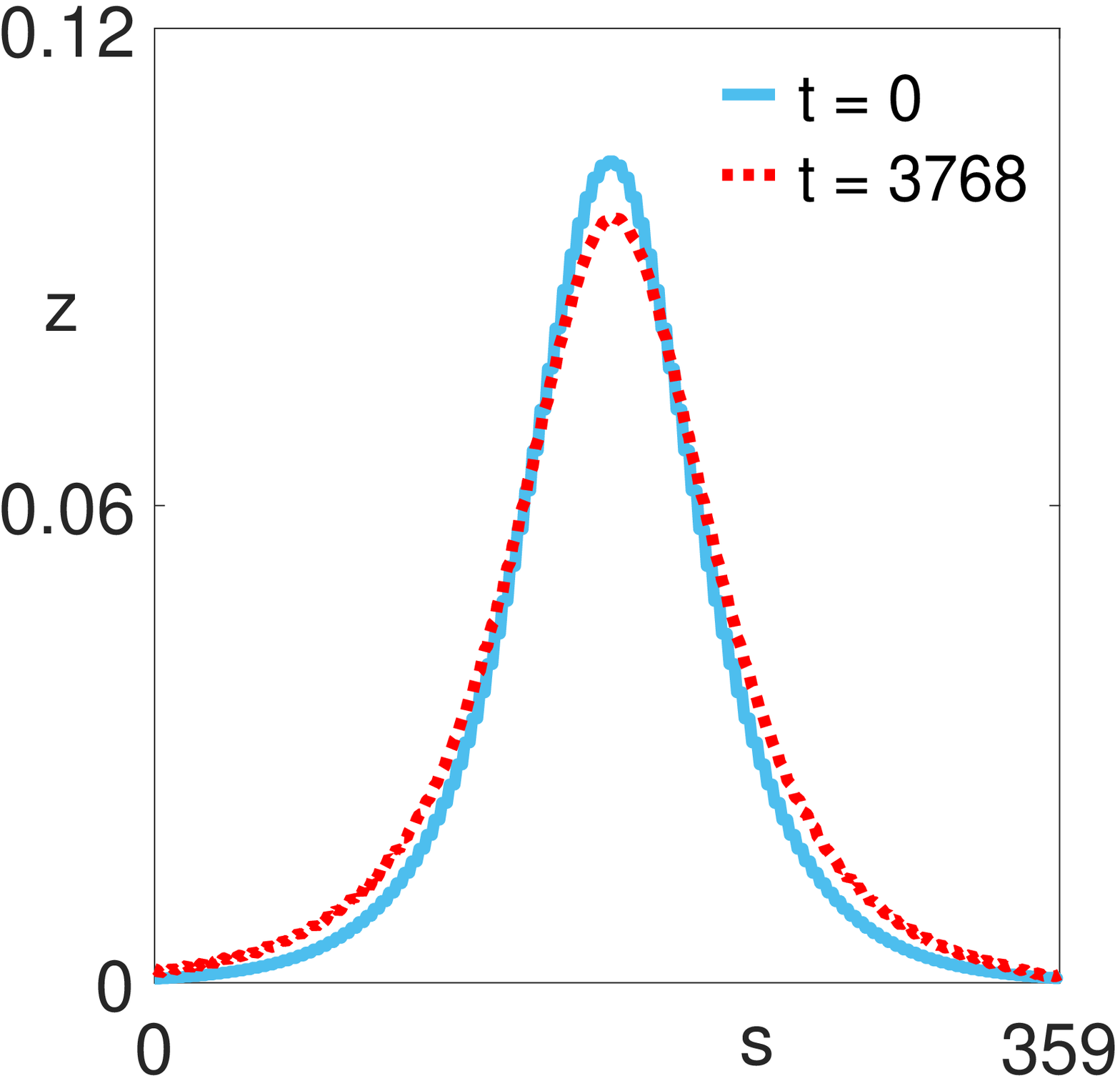}\label{fig:profile-bright}}\hfill
  \subfigure[]{\includegraphics[width=.2\textwidth]{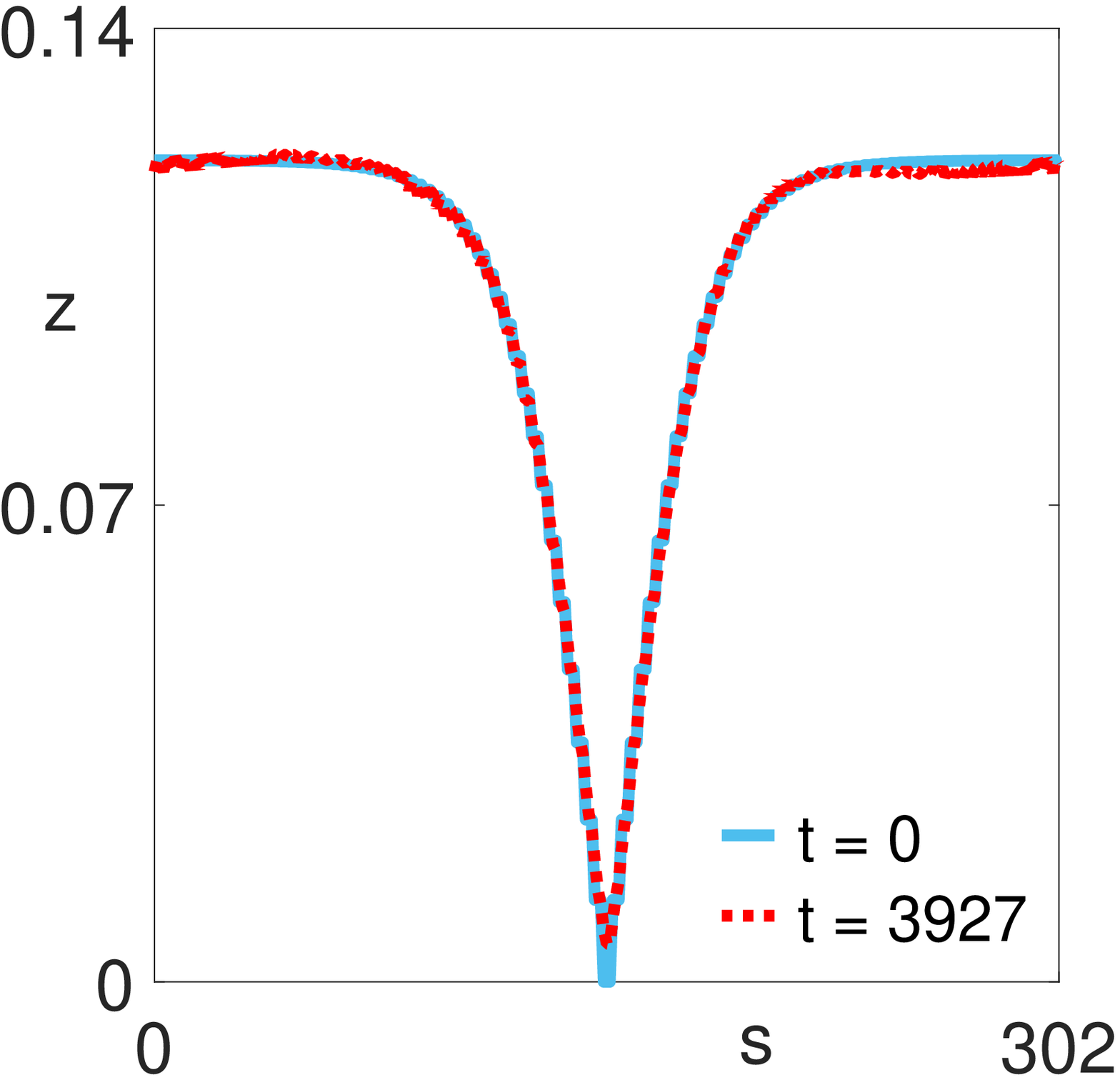}\label{fig:profile-dark}}
     \subfigure[]{\includegraphics[width=.2\textwidth]{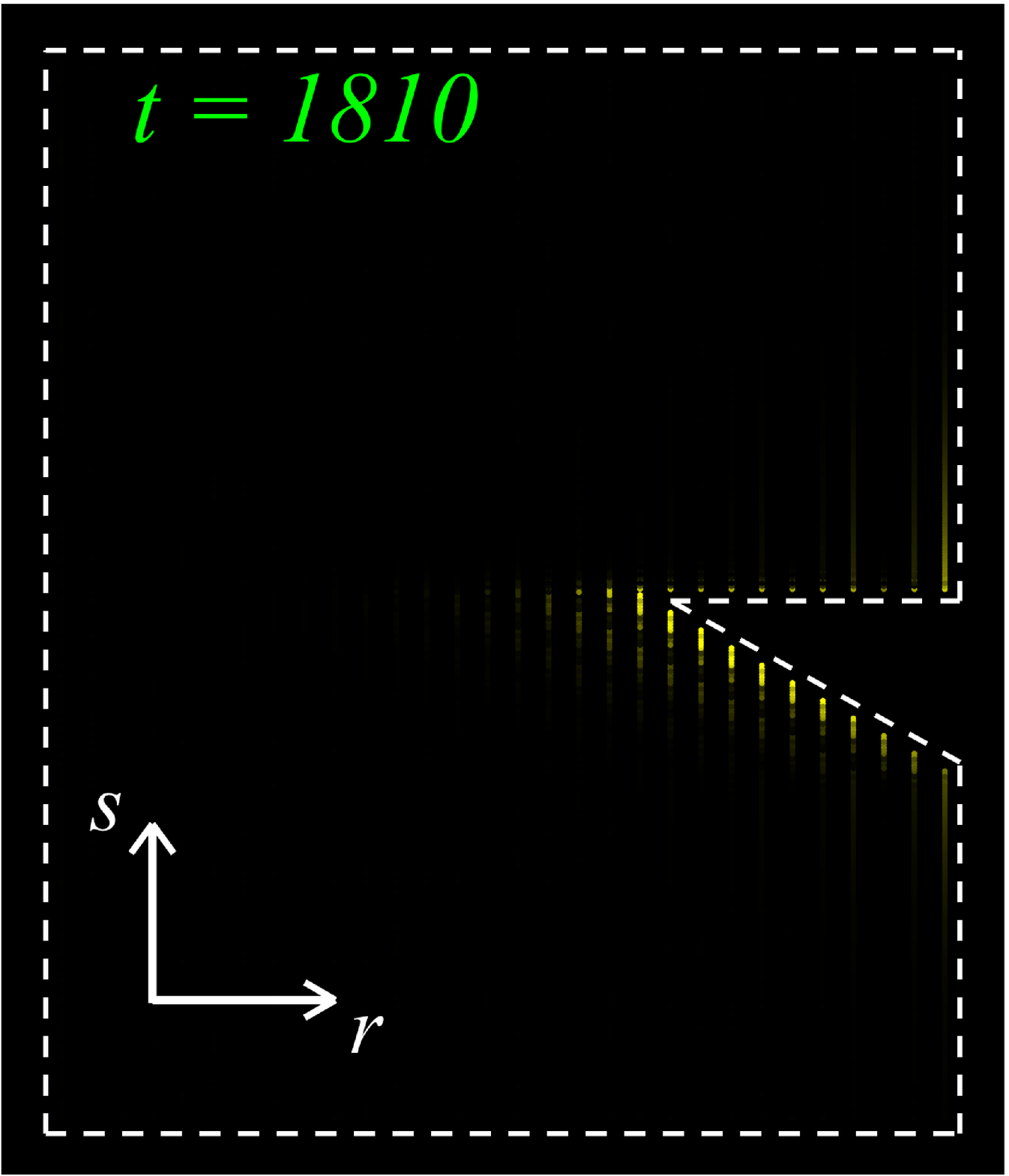}\label{fig:site-missing-triangle}}\hfill
     \subfigure[]{\includegraphics[width=.2\textwidth]{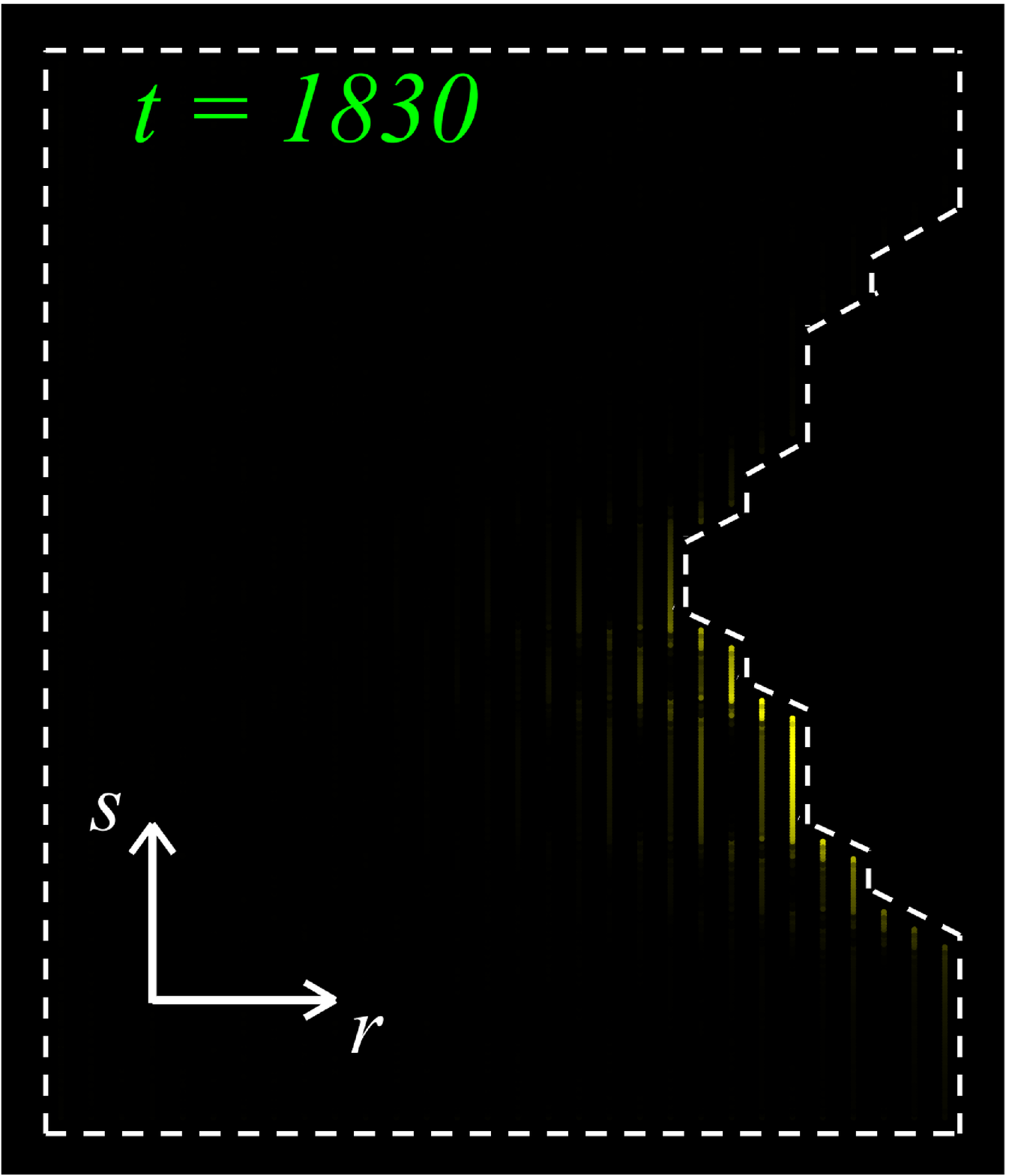}\label{fig:site-missing-cantor}}
    \caption{(Color online) Propagation of bright and dark edge solitons. The 2D domain consists of $N_r$ sites in the $r$-direction and $N_S$ unit cells in the $S$-direction. (a,c,e,f) A bright edge soliton on a rectangular lattice of size $N_r\times N_S=30\times120$. The carrier wavenumber and frequency are $(k_0,\alpha_0)=(2.827,12.506)$, which leads to soliton amplitude $\Lambda=1.376$ in Eq.~(\ref{eq:brightsoliton-envelope}). (b,d) A dark edge soliton on a rectangular lattice of size $N_r\times N_S=30\times101$ with periodic boundary conditions on the top and bottom edges. The carrier wavenumber and frequency are $(k_0,\alpha_0)=(2.146,13.616)$, which leads to soliton amplitude $\Lambda=1.705i$ in Eq.~(\ref{eq:darksoliton-envelope}). Panels (a,b) show space-time plots along the edges, where $L$, $T$, $R$ and $B$ correspond respectively to left, top, right, and bottom edges. Panels (c,d) show the profiles along the left edge initially (blue-solid) and after a long time (red-dashed). Panels (e,f) show snapshots of a bright edge soliton on a rectangular lattice with missing sites on the right edge that form a sharp turn (e) and a leveled structure (f); the white-dashed line is the border of the lattice. Video simulations of (a,c), (b,d), (e), and (f) can be seen in Supplementary Material \cite{Supplementary-Material}.}
    \label{fig:space-time-plots}
\end{figure}

First, we consider the focusing case where Eq.~\eqref{eq:NLS} admits a two-parameter family of bright solitons \cite{Ablowitz-Prinari-2008}, where the two parameters can be taken as $k_0$ and $\epsilon$. The bright soliton is a classical solution of the 1D NLS equation that can be obtained directly from the inverse scattering transform \cite{Shabat-Zakharov-1972}. The focusing condition implies that bright solitons exist only for a finite interval of $k_0$; at each value of $k_0$, fixing the spectral width $\epsilon$ fixes the amplitude of the wave packet. The 2D bright edge soliton can be obtained by considering the initial condition (\ref{eq:general-initial-condition}) with the scalar envelope function
\begin{equation}\label{eq:brightsoliton-envelope}
C_B(S) = \Lambda\mbox{sech}\left(\epsilon(S-S_0)\right),
\end{equation}
where $\Lambda=\sqrt{2 \alpha_0 \alpha''_0 /3\tilde{\sigma}}$ and $S_0$ represents the initial location of the wave packet. According to the theory, since the evolution is asymptotically governed by the 1D NLS equation, the bright edge soliton should persist over a finite time interval. In particular, the edge soliton should maintain its shape better at the theoretically predicted amplitude in comparison to smaller or greater amplitudes, which we analyze in Supplementary Material \cite{Supplementary-Material}. At a large time scale, the asymptotic theory may produce errors that eventually destroy the soliton profile.

In order to express the overall response of a specific site, we create a new variable $z=\sqrt{x^2+y^2}$ which encodes the response of both $x$ and $y$ pendula. The strong transmission of a bright edge soliton can be seen in Fig.~\ref{fig:space-time-bright}; the nonlinear wave traverses the boundaries of the mechanical lattice with almost no energy loss at the corners. This particular edge mode is located in the lower gap of the dispersion relation in Fig.~\ref{fig:dispersion-relation} with carrier wavenumber and frequency $(k_0,\alpha_0)=(2.827,12.506)$. The wave travels clockwise around the domain, completing a full cycle with a period of about $300$. The longtime profile of this edge soliton shows little decay as seen in Fig.~\ref{fig:profile-bright}: at $t\approx2000$ ($t\approx4000$), the soliton has completed about 6 (12) cycles of the domain and the soliton amplitude has only decayed to approximately 96\% (93\%) of its initial theoretically predicted value. This showcases both the topological protection of the traveling wave and the robustness of the soliton profile. 

Akin to bright solitons in focusing NLS, there exists dark soliton solutions to defocusing NLS \cite{Ablowitz-Prinari-2008}, i.e.~the $\alpha''_0\tilde{\sigma}/\alpha_0 < 0$ case in Eq.~(\ref{eq:NLS}). Whereas a bright soliton is a localized rise of energy on a zero background, a dark soliton is a localized dip in energy on a non-zero background. The initial condition for a dark edge soliton is then given as Eq.~(\ref{eq:general-initial-condition}) with the envelope function
\begin{equation}\label{eq:darksoliton-envelope}
C_D(S)=\Lambda \tanh(\epsilon(S-S_0)).
\end{equation}
To facilitate initialization of the non-zero background, we switch over to a rectangular domain that is periodic on the top and bottom edges. Fig.~\ref{fig:space-time-dark} shows a dark edge soliton traveling on such a domain with little energy loss over a long time interval. As shown in Fig.~\ref{fig:profile-dark}, at $t\approx4000$ the carrier wave has not decayed at all, and the dip in energy has only decayed to 96\%. In experiments, any waveform such as Eqs.~(\ref{eq:brightsoliton-envelope}--\ref{eq:darksoliton-envelope}) can be generated at a single site and propagated around the finite domain \cite{Susstrunk-Huber-2015}.

So far we have considered a simple rectangular boundary in order to showcase the ability of the edge solitons to traverse the corners of the domain with ease; however we need not be limited to such trivial boundaries. In fact, such solitons will travel around a domain with any deformation or structure built into it, with little energy loss as before. For instance let us carve a sharp turn into the right edge of the lattice (the edge opposite the initial condition) and evolve the system using the same bright edge soliton initial condition as in Fig.~\ref{fig:space-time-plots}(a,c). Figure \ref{fig:site-missing-triangle} shows a snapshot of the 2D domain at $t\approx1800$, where we see that the soliton traverses the deformity as if it were a simple edge. The same occurs for rough boundaries with finer length scales, such as a Cantor-like function carved into the same edge, as seen in Fig. \ref{fig:site-missing-cantor}.

\begin{figure}[t]
  \centering
  \subfigure[]{\includegraphics[width=.23\textwidth]{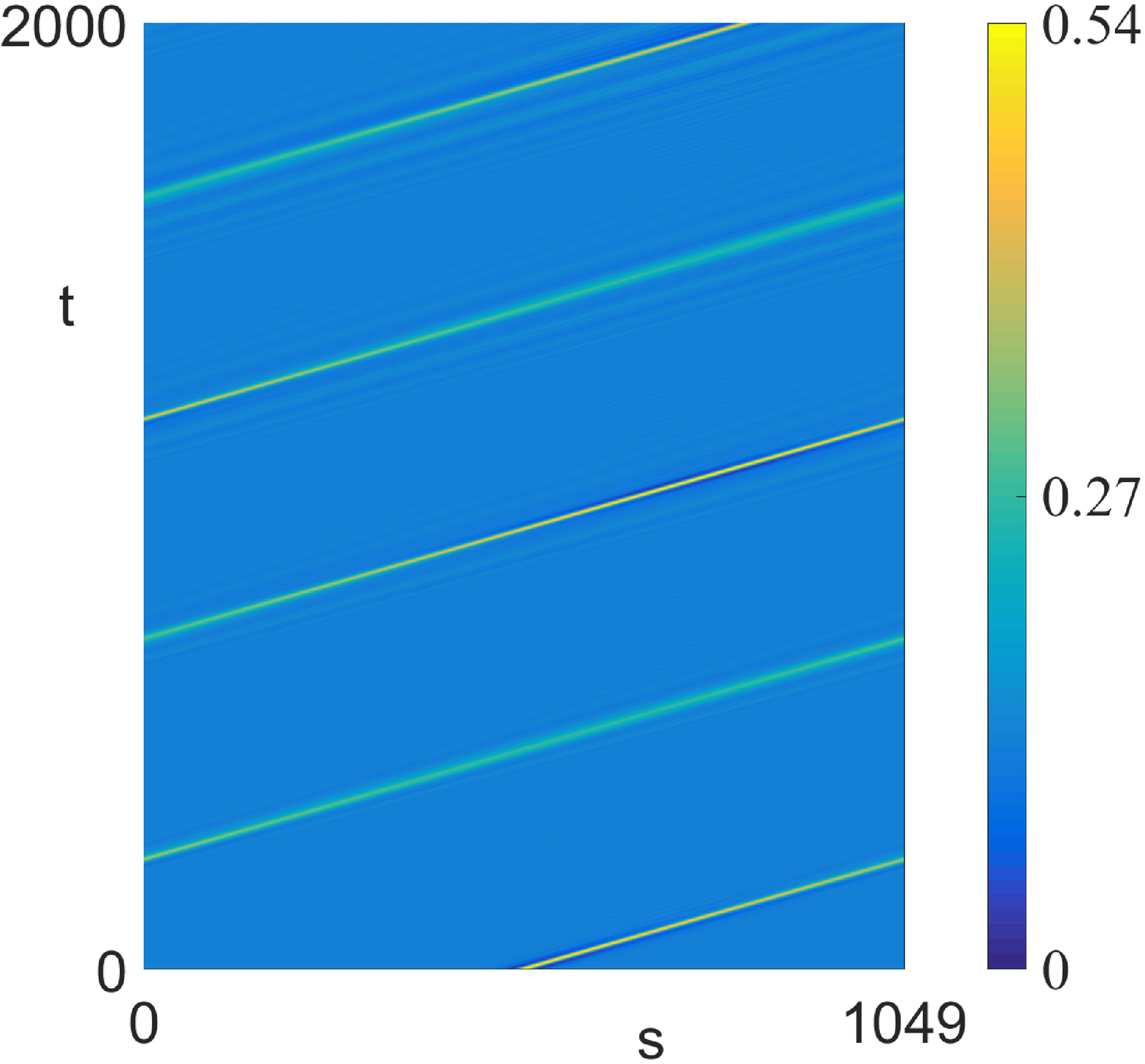}\label{fig:KM}}\hfill
  \subfigure[]{\includegraphics[width=.23\textwidth]{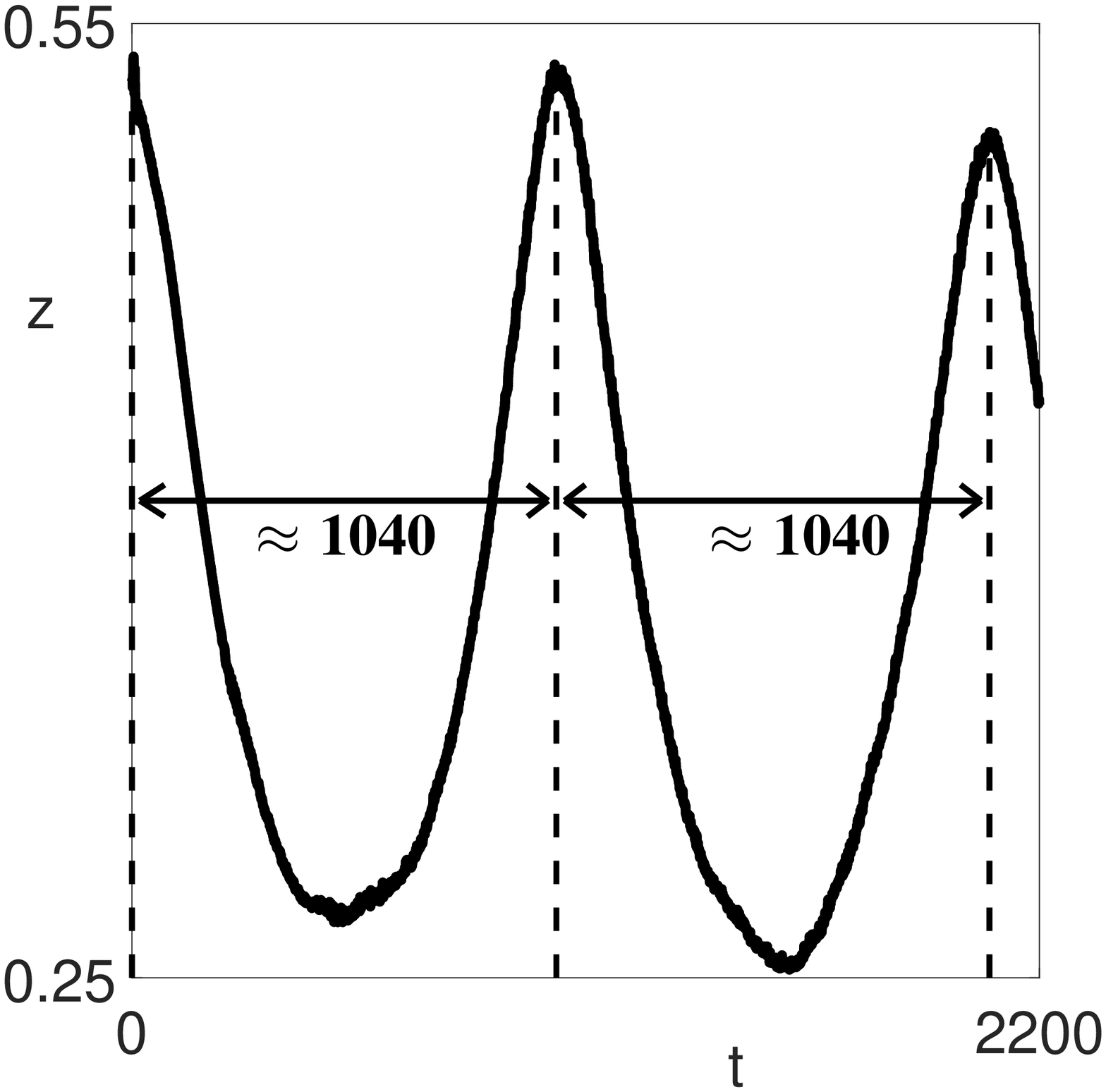}\label{fig:KM-AMP}}
    \subfigure[]{\includegraphics[width=.23\textwidth]{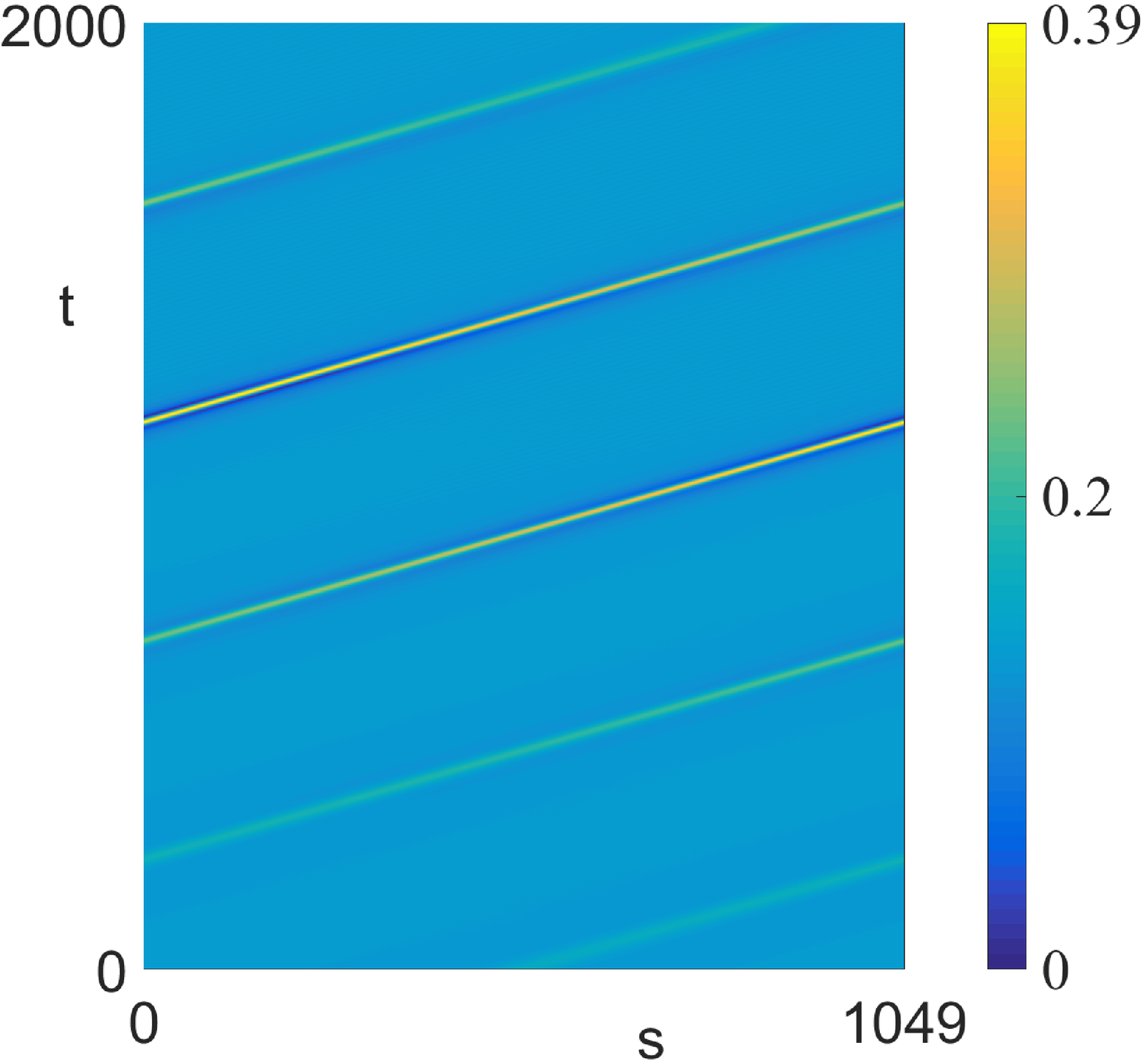}\label{fig:peregrine}}\hfill
  \subfigure[]{\includegraphics[width=.23\textwidth]{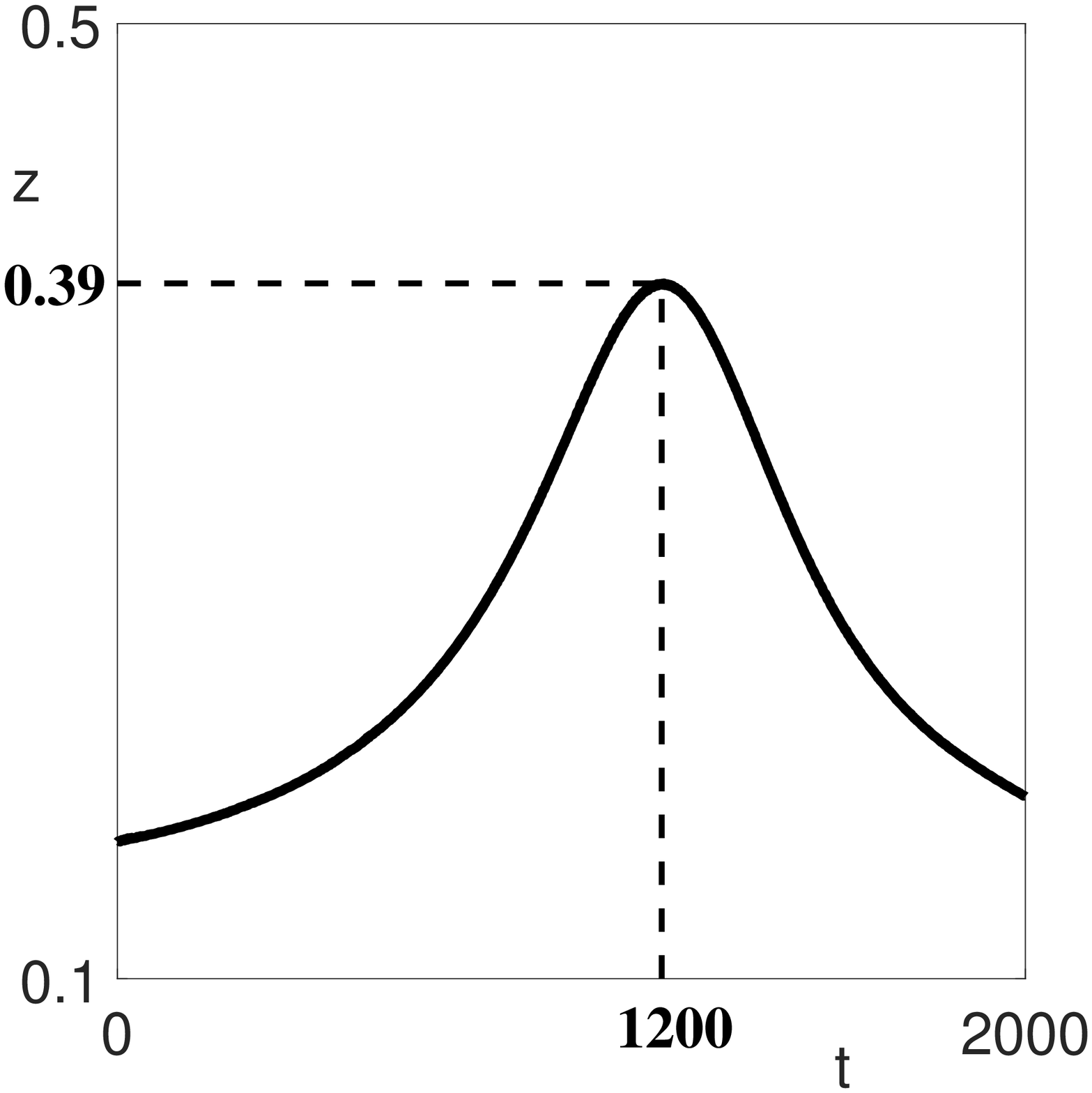}\label{fig:peregrine-AMP}}
    \caption{(Color online) Propagation of rogue edge waves. (a,b) A Kuznetsov-Ma edge soliton described by Eq.~(\ref{eq:KMsoliton-envelope}) with modulation parameter $\phi=1$. (c,d) A Peregrine edge soliton described by Eq.~(\ref{eq:peregrine-envelope}) that peaks at $t\approx1200$. For both, the carrier wavenumber and frequency are $(k_0,\alpha_0)=(2.334,12.113)$, and the rectangular domain is $N_r\times N_S=9 \times 350$ with periodic boundary conditions on the top and bottom edges. Panels (a,c) show space-time plots along the left edge. Panels (b,d) show time evolution of the peak amplitude.}
\end{figure}

The above numerical experiments illustrate the general principle that in mechanical TIs, weakly nonlinear edge solitons inherits the topological protection of linear edge modes. This is true for photonic TIs as well \cite{Ablowitz-Curtis-Ma-2014,Ablowitz-Ma-2015}, and may be understood by suitably generalizing the prior asymptotic analysis. For a finite 2D domain possibly with irregular boundaries and other disorders, there exists a discrete set of 2D topologically protected edge states, which are computable by direct diagonalization. Based on these states, one can construct an edge mode using a similar ansatz to Eq.~(\ref{eq:ansatz}), so long as the domain size is large enough to accommodate the edge mode. The resulting envelope equation is postulated to be the classical 1D NLS equation on a finite periodic domain. Now the phenomenology of topological protection, which manifests itself as the unidirectional vortex motion of a 2D edge soliton, can be simply identified as the unidirectional periodic motion of a classical 1D NLS soliton.

Finally, we explore certain temporally or spatially periodic solutions to the classical 1D NLS equation, commonly known as rogue waves as they have recently proven to be very promising contenders for modeling waves with abnormally large heights \cite{Kharif-Pelinovsky-2003,Chabcoub-etal-2011,Dysthe-Trulsen-1999,BlKo09_PRA}. We first consider the Kuznetsov-Ma (K-M) soliton, which was first derived by Ma \cite{Ma-1979} as a breathing wavepacket in time embedded in a plane wave solution. To generate a K-M edge soliton, we replace the envelope function in the initial condition (\ref{eq:general-initial-condition}) by the K-M representation seen in \cite{Dysthe-Trulsen-1999}, i.e.
\begin{equation}\label{eq:KMsoliton-envelope}
C_{KM}(S;\phi)=\Lambda\left[\frac{\cos(2i\phi)-\cosh\phi\cosh\bar{S}}{1-\cosh\phi\cosh\bar{S}}\right],
\end{equation}
where $\bar{S}=P\epsilon(S-S_0)$, $P=2\sinh\phi$, and $\phi$ is the real modulation parameter defining the breathing period. Figure \ref{fig:KM} shows the appearance of a K-M edge soliton with $\phi=1$ and the carrier wavenumber and frequency $(k_0,\alpha_0)=(2.334,12.113)$. Figure \ref{fig:KM-AMP} shows a plot of the maximum amplitude of the soliton over time, with the breathing period found to agree with the theory.

We can also briefly explore the famous Peregrine soliton \cite{Peregrine-1983} which is just a limiting case ($\phi\rightarrow 0$) of the K-M soliton found previously. For visualization purposes, we require the maximum amplitude to appear at some $t>0$, say $t_P=1200$. Therefore we consider the initial condition (\ref{eq:general-initial-condition}) with the envelope function given by  
\begin{equation}\label{eq:peregrine-envelope}
C_P(S) =\Lambda \left[ 1- \frac{4(1-2i\tilde{t}_P)}{1+4\epsilon^2(S-S_0)^2+4\tilde{t}_P^2} \right] e^{i\tilde{\theta}},
\end{equation}
where $\tilde{\theta}=\alpha_0 t_P-\tilde{t}_P$ and $\tilde{t}_P=\epsilon^2\alpha_0''t_P$. Figure \ref{fig:peregrine} shows the appearance of the Peregrine edge soliton with the same carrier wavenumber and frequency as in the K-M case. Figure \ref{fig:peregrine-AMP} shows a plot of the maximum amplitude of the soliton over time, with the peak indeed appearing at $t\approx t_P$. Note that since the carrier wave belongs to the topologically nontrivial regime, the propagation of these rogue edge waves in the mechanical lattice is topologically protected, just like the traveling edge solitons presented earlier. See Supplementary Material \cite{Supplementary-Material} for other rogue waves such as the Akhmediev breather \cite{Akhmediev-etal-1987} and other traveling waves such as gray solitons \cite{Ablowitz-Prinari-2008}. See also \cite{Charalampidis-etal-2018} for the existence of rogue waves in the FPUT lattice and granular crystals through reduction to NLS.

In summary, we have realized edge solitons theoretically and numerically in a 2D nonlinear mechanical topological insulator exhibiting the QSHE. The asymptotic analysis shows that these nonlinear edge solitons are governed by the classical 1D NLS equation and are topologically protected in the same way as linear edge waves. The derivation can be readily generalized to other mechanical TIs with discrete elements \cite{Pal-etal-2016,SBOPTPC17_NJP}, possibly with dissipation \cite{Xiong-etal-2016} and forcing \cite{Pal-etal-2018} included. Further extensions may include TIs in continuous media that can exhibit significant nonlinearity, such as recent proposals based on magnetic solitons \cite{KT17_PRL} and water waves \cite{WWM18_NJP}. The existence of TPES in both photonic and phononic TIs may have significant impacts on practical applications such as optical and acoustic delay lines \cite{Hafezi-etal-2011,ZhTi_PRAppl18} and robust manipulation of light and sound \cite{Cheng-etal-2016,YvFl17_NJP}.

Y.M. acknowledges support from a Vice Chancellor's Research Fellowship at Northumbria University.

\newpage
\clearpage

\renewcommand{\thefigure}{S\arabic{figure}}

\setcounter{figure}{0}

\begin{center}
\textbf{\large{Supplementary Material:}} \\
\large{Edge Solitons in a Nonlinear Mechanical Topological Insulator} \\
\small{David D. J. M. Snee and Yi-Ping Ma} 
\end{center}
The Supplementary material here is organized into four sections and accompanying films for the bright and dark soliton simulations. The first section entails the equations of motion for the mechanical lattice which is adopted from S{\"u}sstrunk and Huber \cite{Susstrunk-Huber-2015} and subsequently adapted to include the cubic nonlinearity inherent to pendula. In the second section we look closely at the bright and dark solitons and show that the theoretically predicted NLS amplitude derived from the asymptotic analysis is the most efficient amplitude for the evolution of solitons in the system. Next we look at the interaction of two bright solitons initialized on the same lattice and qualitatively show the phase shift that occurs in such a collision. Finally we verify the nonlinear theory further by exploring more exotic solutions to the classical 1D NLS equation, such as gray solitons and other rogue wave solutions, and realize them in the mechanical topological insulator.       

\begin{center}
\textsc{EQUATIONS OF MOTION OF COUPLED MECHANICAL OSCILLATORS} \\
\end{center}

The 6 nonlinear equations of motion describing this system of coupled oscillating pendula for $x_{r,S}^{(0)}$, $y_{r,S}^{(0)}$, $x_{r,S}^{(1)}$, $y_{r,S}^{(1)}$, $x_{r,S}^{(2)}$, and $y_{r,S}^{(2)}$ are given as follows,

\begin{myequation}\label{eq:x0}
\begin{split}
\ddot{x}^{(0)}_{r,S} & =-(\omega_0^2+Af)x^{(0)}_{r,S}+\sigma (x^{(0)}_{r,S})^3 \\ & +f(x^{(0)}_{r+1,S}+x^{(0)}_{r-1,S}+x^{(1)}_{r,S}+x^{(2)}_{r,S-1}),
\end{split}
\end{myequation}
\begin{myequation} \label{eq:y0}
\begin{split}
\ddot{y}^{(0)}_{r,S} & =-(\omega_0^2+Af)y^{(0)}_{r,S}+\sigma (y^{(0)}_{r,S})^3 \\ &+f(y^{(0)}_{r+1,S}+y^{(0)}_{r-1,S}+y^{(1)}_{r,S}+y^{(2)}_{r,S-1}),
\end{split}
\end{myequation}
\begin{myequation} \label{eq:x1}
\begin{split}
\ddot{x}^{(1)}_{r,S} & =-(\omega_0^2+Af)x^{(1)}_{r,S}+\sigma (x^{(1)}_{r,S})^3 \\ & +f(x^{(0)}_{r,S}+x^{(2)}_{r,S})-\frac{f}{2}(x^{(1)}_{r+1,S}+x^{(1)}_{r-1,S}) \\ & +\frac{\sqrt{3}f}{2}(y^{(1)}_{r+1,S}-y^{(1)}_{r-1,S}),
\end{split}
\end{myequation}
\begin{myequation} \label{eq:y1}
\begin{split}
\ddot{y}^{(1)}_{r,S} & =-(\omega_0^2+Af)y^{(1)}_{r,S}+\sigma (y^{(1)}_{r,S})^3 \\ & +f(y^{(0)}_{r,S}+y^{(2)}_{r,S})-\frac{f}{2}(y^{(1)}_{r+1,S}+y^{(1)}_{r-1,S}) \\ & +\frac{\sqrt{3}f}{2}(-x^{(1)}_{r+1,S}+x^{(1)}_{r-1,S}),
\end{split}
\end{myequation}
\begin{myequation} \label{eq:x2}
\begin{split}
\ddot{x}^{(2)}_{r,S} & =-(\omega_0^2+Af)x^{(2)}_{r,S}+\sigma (x^{(2)}_{r,S})^3 \\ & +f(x^{(0)}_{r,S+1}+x^{(1)}_{r,S})-\frac{f}{2}(x^{(2)}_{r+1,S}+x^{(2)}_{r-1,S}) \\ & +\frac{\sqrt{3}f}{2}(-y^{(2)}_{r+1,S}+y^{(2)}_{r-1,S}),
\end{split}
\end{myequation}
\begin{myequation} \label{eq:y2}
\begin{split}
\ddot{y}^{(2)}_{r,S} & =-(\omega_0^2+Af)y^{(2)}_{r,S}+\sigma (y^{(2)}_{r,S})^3 \\ & +f(y^{(0)}_{r,S+1}+y^{(1)}_{r,S})-\frac{f}{2}(y^{(2)}_{r+1,S}+y^{(2)}_{r-1,S}) \\ & +\frac{\sqrt{3}f}{2}(x^{(2)}_{r+1,S}-x^{(2)}_{r-1,S}).
\end{split}
\end{myequation}

The constants of the system $A$, $f$, $\omega_0$, and $\sigma$ describe the forces due to the springs and nonlinear pendula. The restoring acceleration of the springs is encoded by terms proportional to $A=3+\sqrt{3}$, whereas the springs themselves are described by $f=M/J$; $M$ is the torque constant and $J$ is the moment of inertia. Consistent with \cite{Susstrunk-Huber-2015} we choose the length of the pendula to be 500mm carrying a mass of 500g, which gives $\omega_0\approx3\pi/2$ and $f\approx4.16\pi^2$. The nonlinear coefficient $\sigma$ is found by expanding the restoring force of the pendula, i.e. $-\omega_0^2\sin(\boldsymbol{X}_{r,S})=-\omega_0^2\boldsymbol{X}_{r,S}+\frac{\omega_0^2}{6}\boldsymbol{X}_{r,S}^3$, which gives $\sigma=\frac{\omega_0^2}{6}$. The spacing between lattice sites is given as 135mm in $r$ and 120mm in $s$ and the pendula are hinged such that they are only free to swing in the $s$-direction.

\begin{figure}[ht]
  \centering
  \subfigure[]{\includegraphics[width=.23\textwidth]{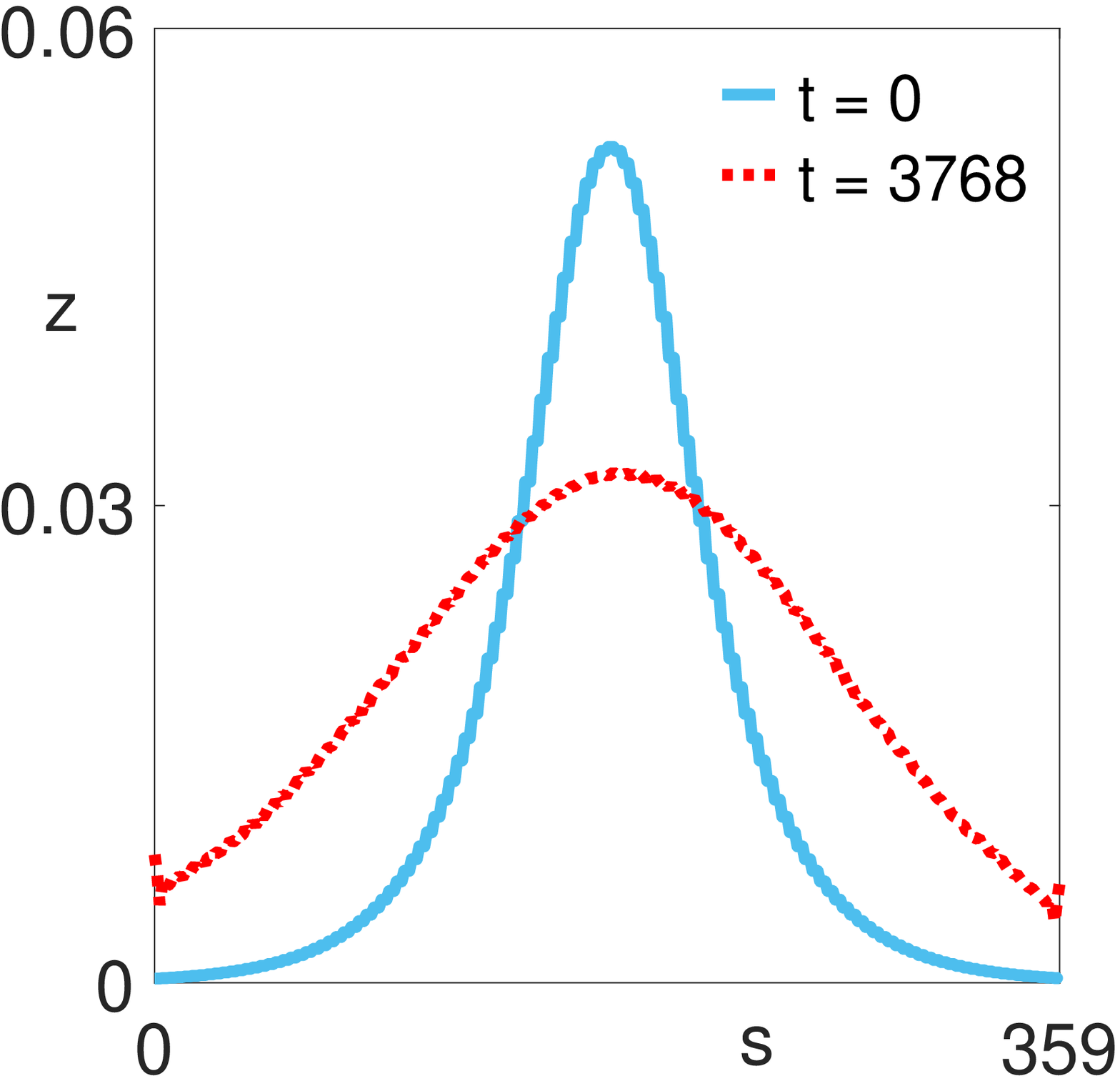}\label{fig:bright-profile-007}}\hfill
  \subfigure[]{\includegraphics[width=.23\textwidth]{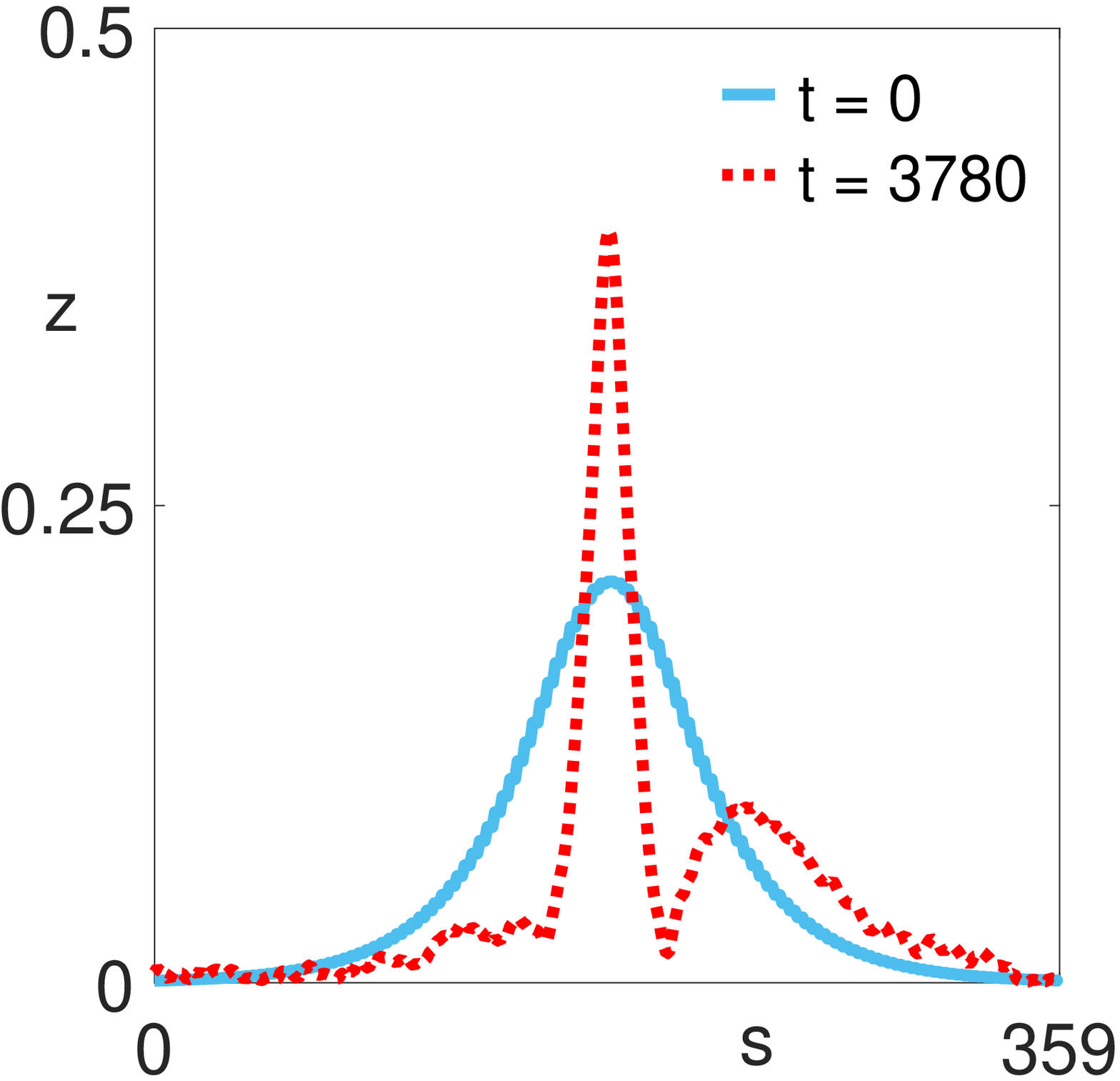}\label{fig:bright-profile-028}}
    \subfigure[]{\includegraphics[width=.23\textwidth]{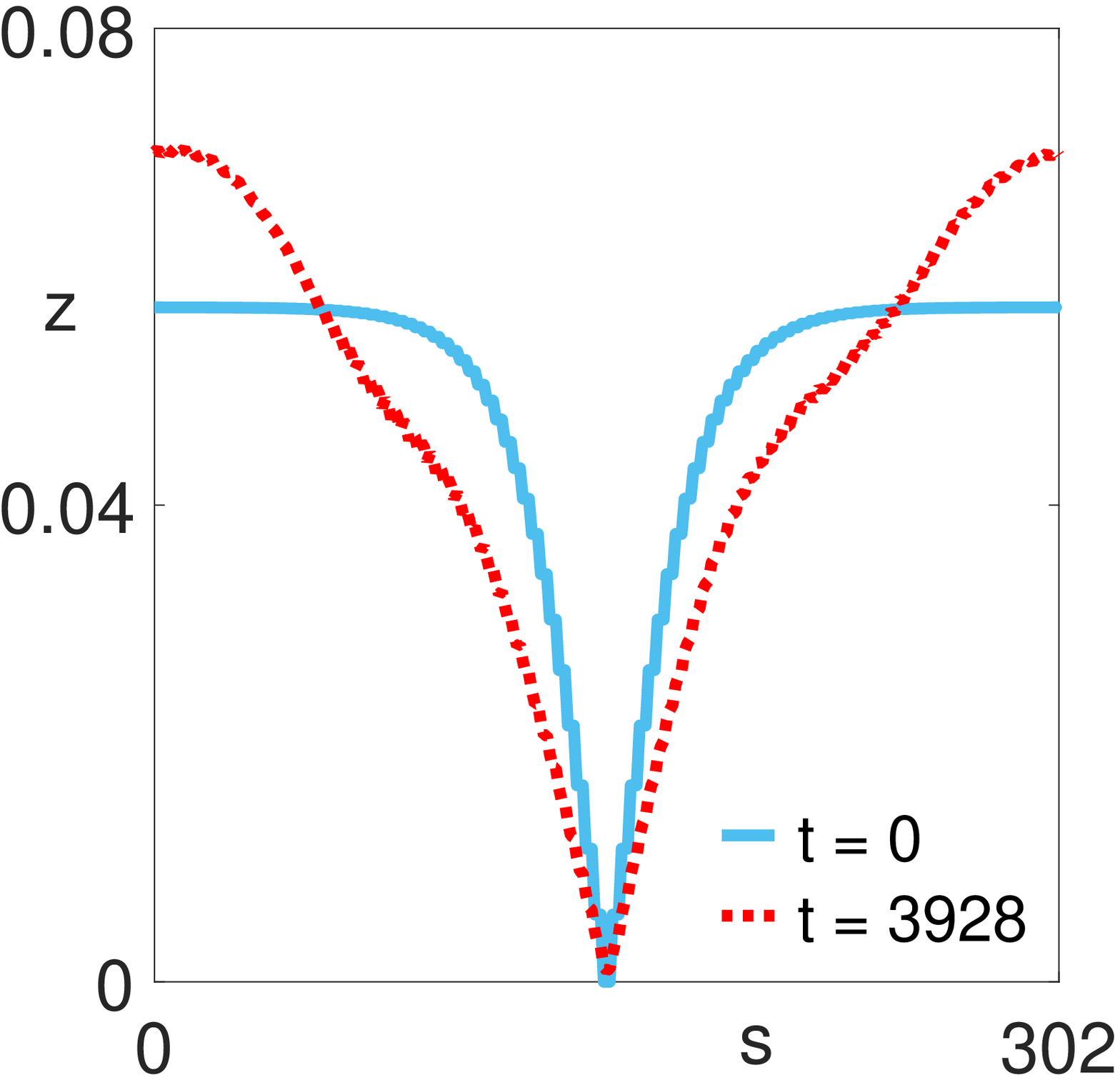}\label{fig:dark-profile-008}}\hfill
  \subfigure[]{\includegraphics[width=.23\textwidth]{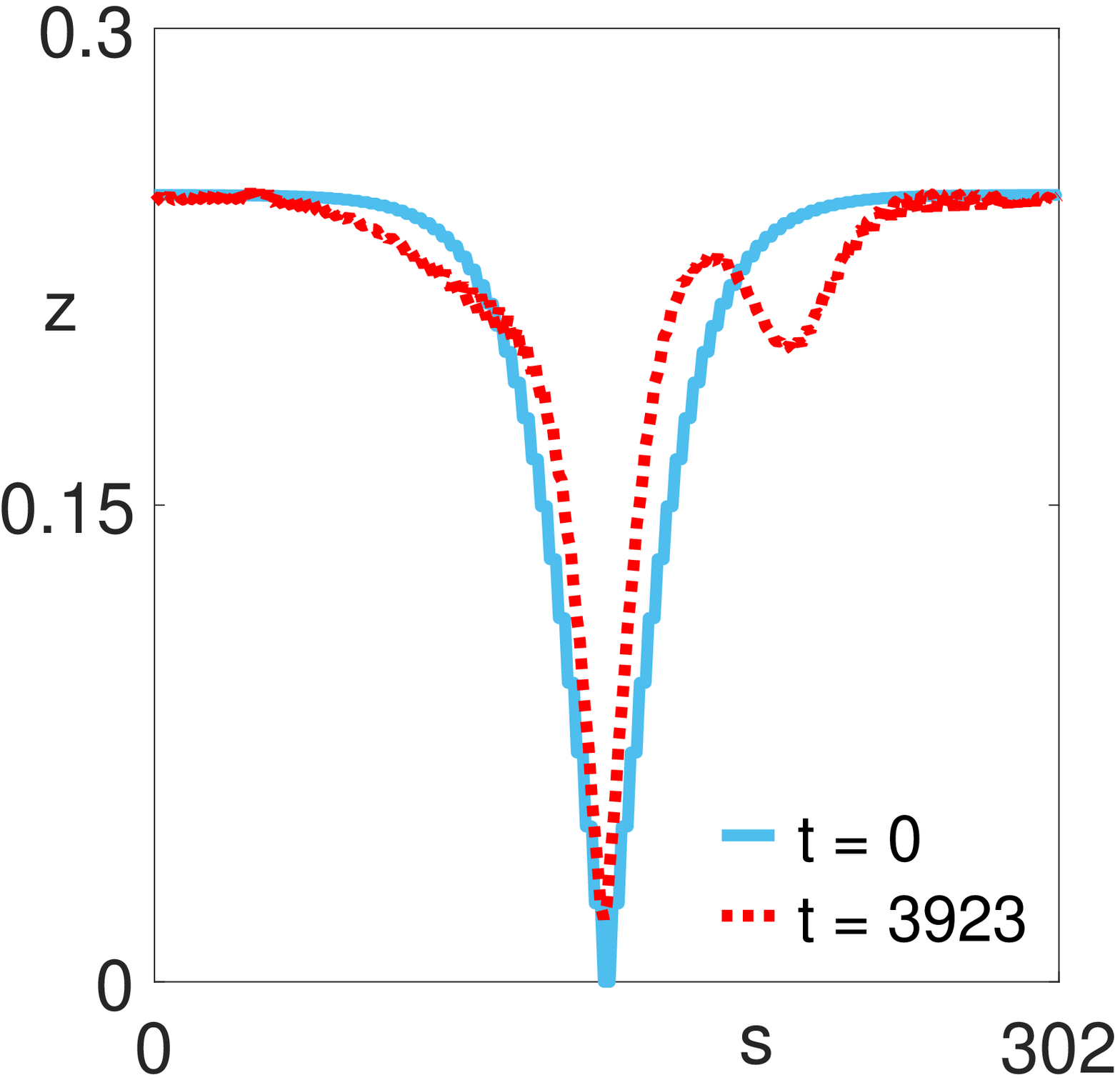}\label{fig:dark-profile-035}}
    \caption{(Color online) Transmission of bright solitons (a,b) with initial amplitudes (a) $\Lambda=0.7$; (b) $\Lambda=2.8$, wavenumber $k_0=2.827$, and corresponding frequency $\alpha_0=12.506$. Transmission of dark solitons (c,d) with initial amplitudes (c) $\Lambda=0.8i$; (d) $\Lambda=3.5i$, wavenumber $k_0=2.146$, and frequency $\alpha_0=13.616$. Profiles are taken initially (blue-solid) and after a long time interval (red-dashed). The amplitudes of the solitons are comparable to the theoretically predicted NLS amplitude seen in Fig. \ref{fig:space-time-plots}.}
\end{figure}

\begin{center}
\textsc{THE NLS AMPLITUDE} \\
\end{center}

We will now consider what happens when we force an amplitude other than the one predicted by the asymptotic analysis onto the bright soliton envelope (\ref{eq:brightsoliton-envelope}) and the dark soliton envelope (\ref{eq:darksoliton-envelope}) whilst keeping the other parameters unchanged. 

Firstly, let us take the bright edge soliton in Fig. \ref{fig:space-time-bright}, with wavenumber $k_0=2.827$, frequency $\alpha_0=12.506$, and theoretically predicted NLS amplitude $\Lambda=1.376$, as a case study for the varied amplitudes. Taking a smaller amplitude than the theoretically predicted results in a compelling amount of decay of the bright soliton profile. This is seen in Fig. S\ref{fig:bright-profile-007} where we have considered an amplitude that is approximately half of the predicted, i.e. $\Lambda=0.7$. Here, the peak of the soliton has decayed to 61\% of its original initialized peak long after the initial time ($\approx 4000$ time units), which is a significant decrease in comparison to the 93\% seen in the case of the NLS amplitude (Fig. \ref{fig:profile-bright}). Moreover, the Gaussian profile seems to be flattened and spread out to the point where the tail ends of the bright soliton are being cut off by the corners of the 2D rectangular domain. In contrast to this Fig. S\ref{fig:bright-profile-028} shows us the effect of initializing a larger amplitude on the mechanical system, where we have considered approximately double the theoretically predicted, i.e $\Lambda=2.8$. We see that this has a severe consequence on the structure of the bright soliton and the crest of the initial Gaussian curve has almost doubled in size; the remainder of the soliton has a somewhat random structure to it. This phenomena is down to an overfocusing effect where the energy to the immediate left and right of the initial peak has been redistributed to the peak itself.

The dark soliton can be explored in a similar manner by initializing an amplitude of half and double the theoretically predicted amplitude, seen in Figures S\ref{fig:dark-profile-008} and S\ref{fig:dark-profile-035} respectively. These two profiles are comparable to the dark soliton profile in Fig. \ref{fig:profile-dark} with predicted amplitude $\Lambda=1.705i$, wavenumber $k_0=2.146$, and corresponding frequency $\alpha_0=13.616$. We see once again that the NLS amplitude is the best amplitude in the evolution of the dark soliton. When once considers an amplitude which is less than that of the theoretically predicted amplitude then we see that, despite there being no extra decay in the dip of the energy, the trough of the dark soliton widens and the carrier wave has effectively risen in amplitude. Contrary to this, taking an initial amplitude which is greater than that of the theoretically predicted causes a distortion in the profile itself and produces a second dip in energy from the carrier wave.

Overall, it is clear to see that the amplitude theoretically predicted by the NLS equation (\ref{eq:NLS}) which governs the envelope of these nonlinear traveling waves is truly the best amplitude for the evolution of this MTI. If one considers an initial amplitude other than this then the nonlinear and dispersive terms are no longer balanced causing effects such as overfocusing and severe profile distortion as the soliton evolves in time.  

\begin{center}
\textsc{INTERACTION OF TWO BRIGHT SOLITONS}
\end{center}

\begin{figure}[t]
  \centering
  \subfigure[]{\includegraphics[width=.5\textwidth]{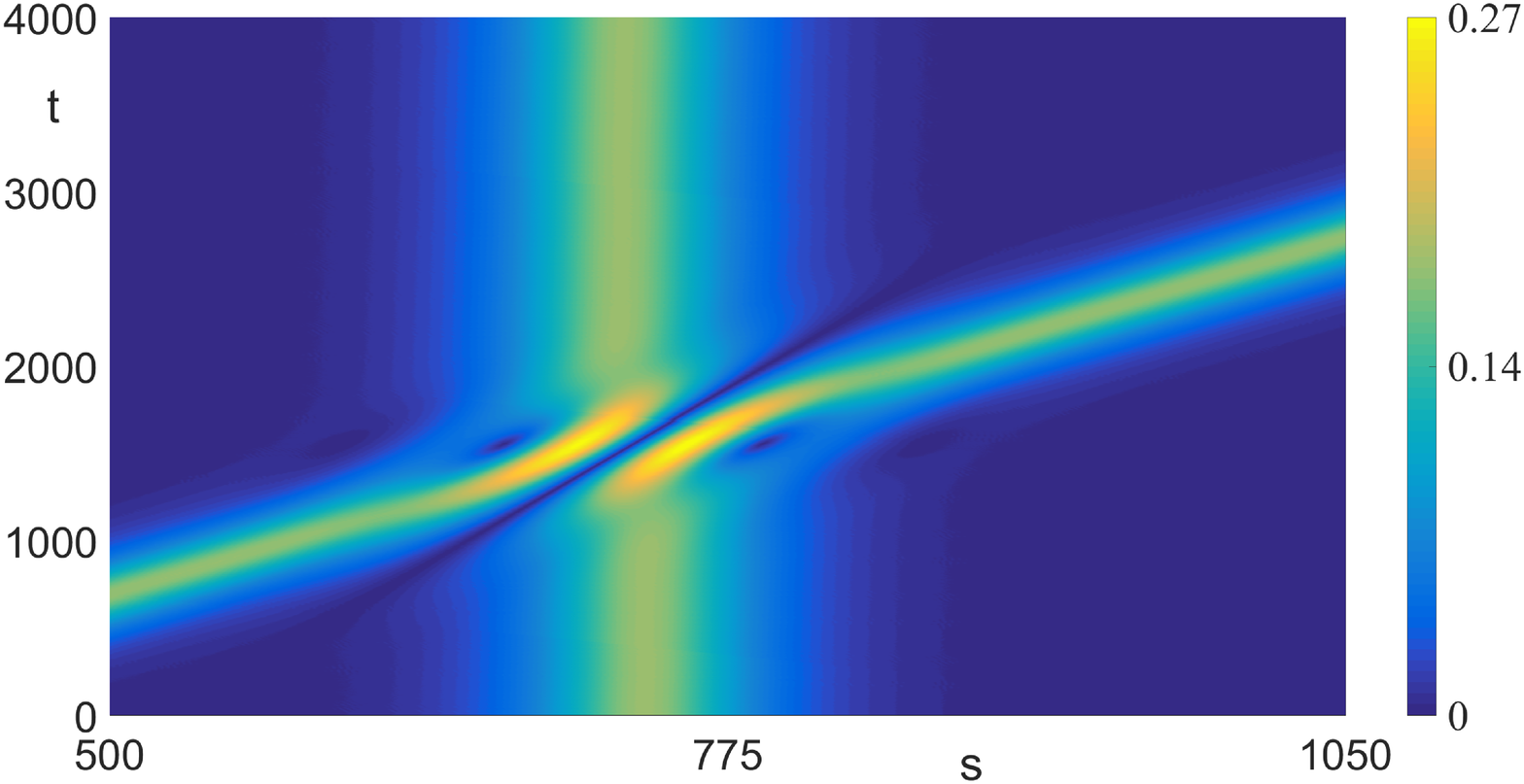}\label{fig:collision}}
    \caption{(Color online) Phase shift of a bright soliton directly after the interaction with another bright soliton. The co-moving soliton has wavenumber $k_0=0.524$, frequency $\alpha_0=11.138$, and initial amplitude $\Lambda=2.345$, whilst the interacting soliton has $k_0=0.785$, $\alpha_0=11.219$, and $\Lambda=2.306$.}
\end{figure}

Let us now consider the case where we initialize two bright solitons on the same edge of the MTI and allow them to collide. For this we move to an $s$-periodic domain and transfer to the co-moving frame of one of the waves so to best visualize the structure of the collision that occurs. Since the waves are nonlinear we expect some nonlinear effects to take place and the interaction should not be a simple linear superposition of the two waves.

Figure. S\ref{fig:collision} shows the interaction of the two bright solitons on a periodic domain of $9 \times 1050$. Here we allow both waves to travel up the periodic edge with one wave having a greater group velocity, thus catching the other wave some time after initializing. The slower soliton has wavenumber $k_0=0.524$ and corresponding frequency $\alpha_0=11.138$, whilst the faster soliton has wavenumber $k_0=0.785$ and frequency $\alpha_0=11.219$. Here we transfer into the co-moving frame of the slower wave. The figure shows that immediately after the collision has occurred, the faster wave has passed through the slower with both structures intact and with the same amplitude and group velocity prior to the interaction. The slower wave has been visibly phase-shifted as a direct consequence of the collision and the interaction center is not a simple linear sum of the two amplitudes as we would expect from nonlinear waves.   

\begin{center}
\textsc{EXOTIC SOLUTIONS IN THE MECHANICAL TOPOLOGICAL INSULATOR} \\
\end{center}

We have already seen particular examples of soliton and rogue wave solutions realized in this mechanical topological insulator. There is however, the existence of many more, what we call exotic, solutions to the classical 1D NLS which theoretically produces an analogue structure in our system. Such solutions include the the gray soliton and Akhmediev breather, and if the envelope of the nonlinear waves is truly governed by the NLS equation then these exotic solutions should be realizable in the MTI.

\begin{figure}[t]
  \centering
  \subfigure[]{\includegraphics[width=.23\textwidth]{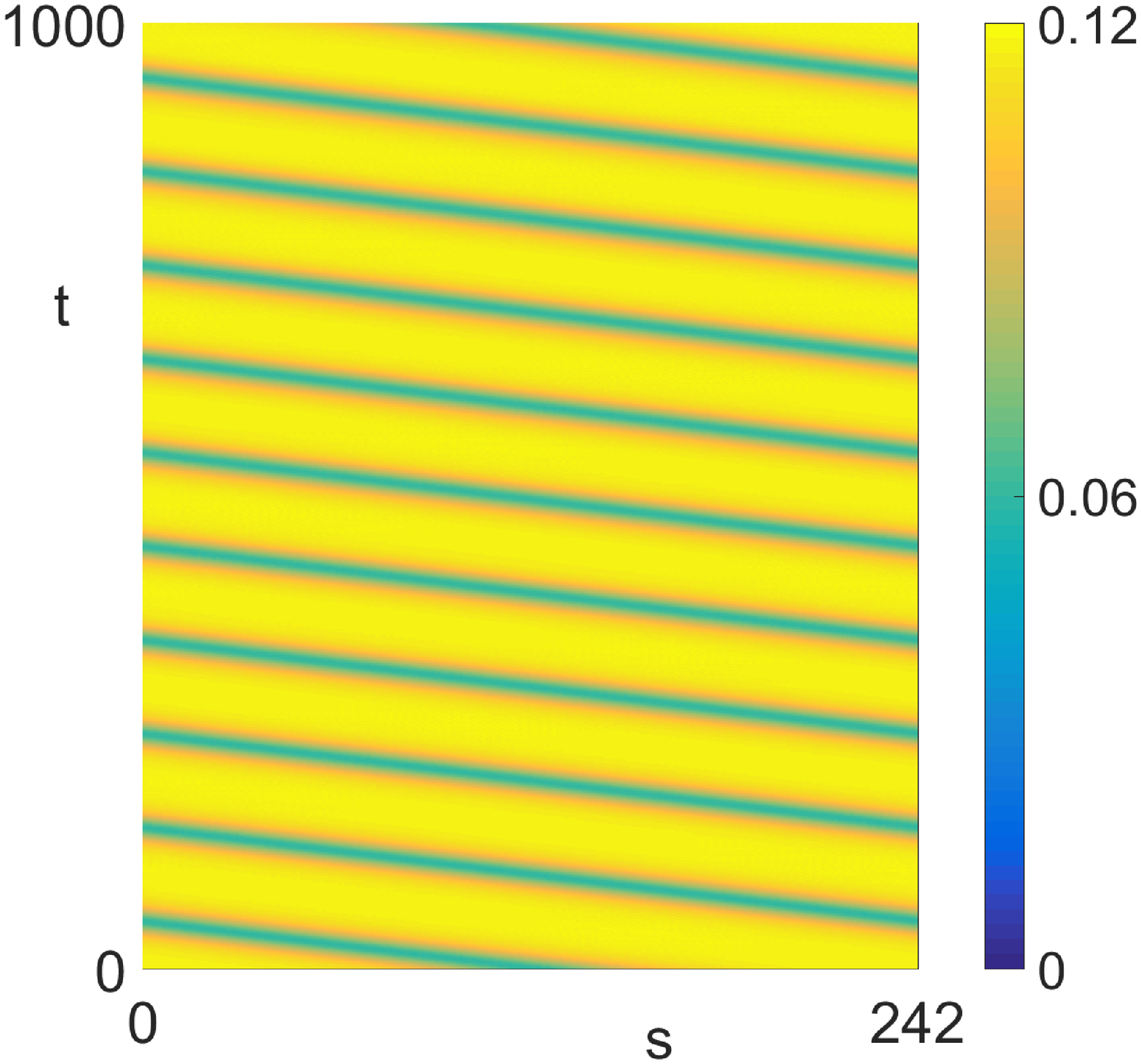}\label{fig:grey-soliton}}\hfill
  \subfigure[]{\includegraphics[width=.23\textwidth]{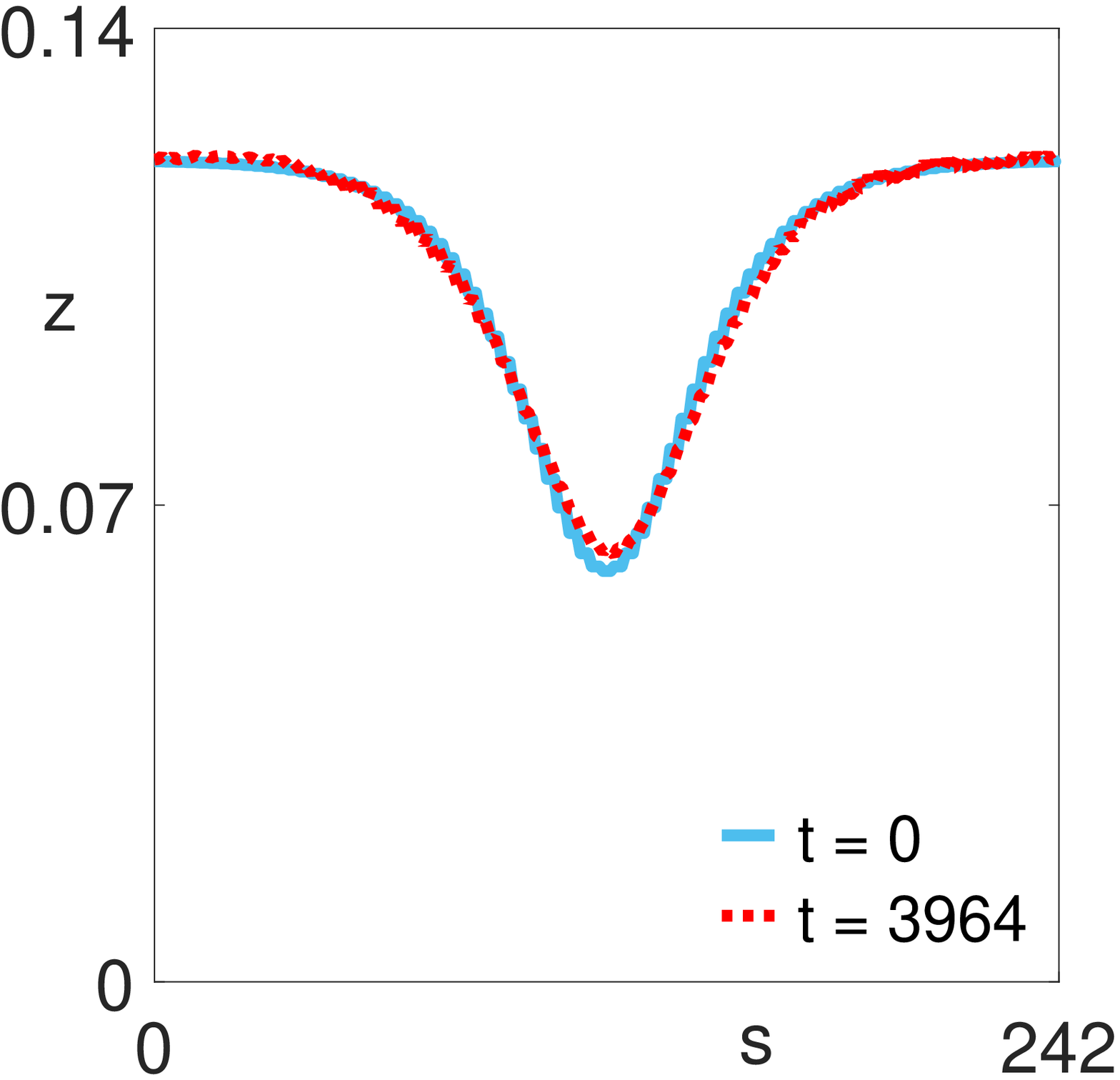}\label{fig:profile-grey}}
    \subfigure[]{\includegraphics[width=.23\textwidth]{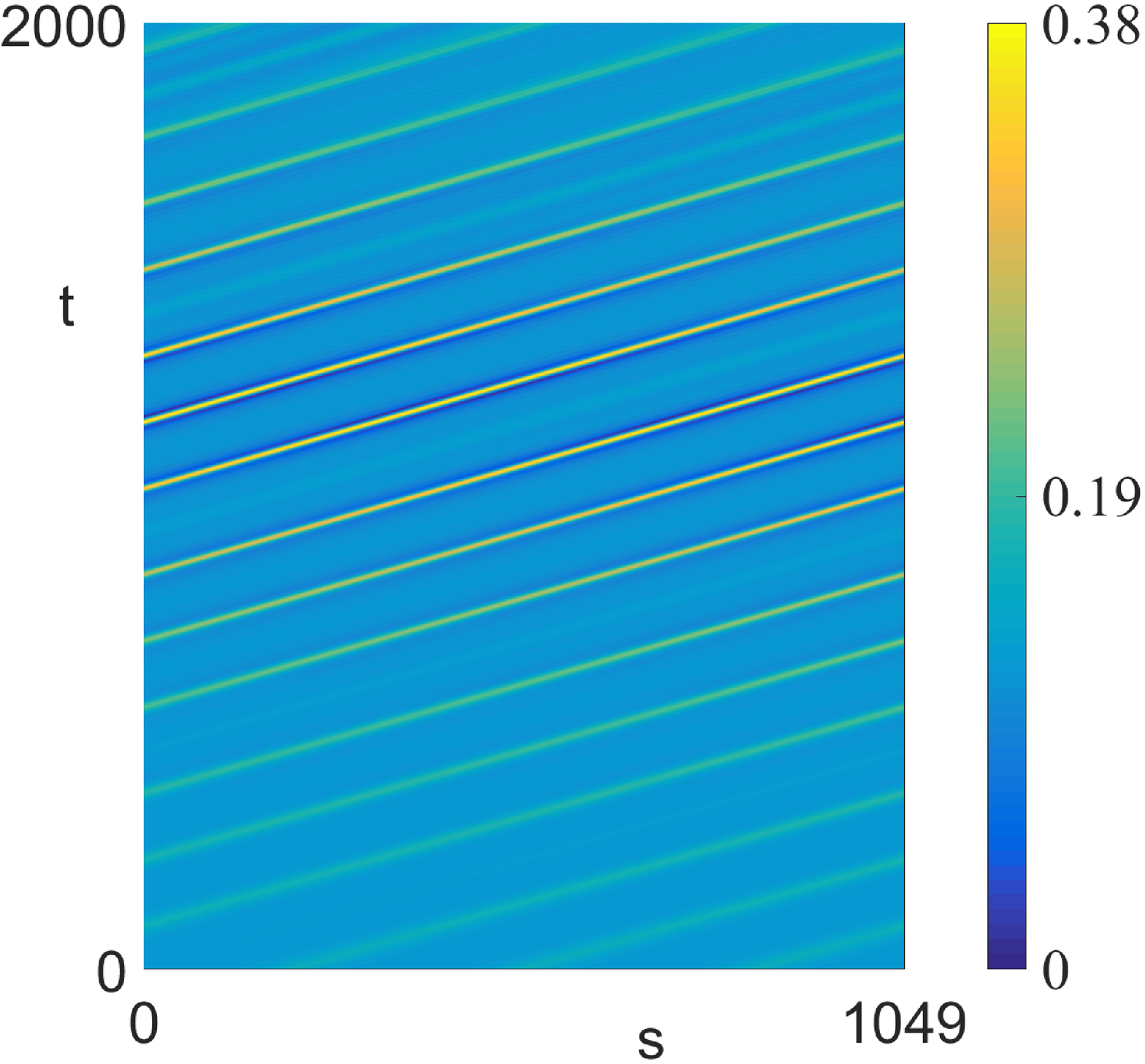}\label{fig:A-breather}}\hfill
  \subfigure[]{\includegraphics[width=.23\textwidth]{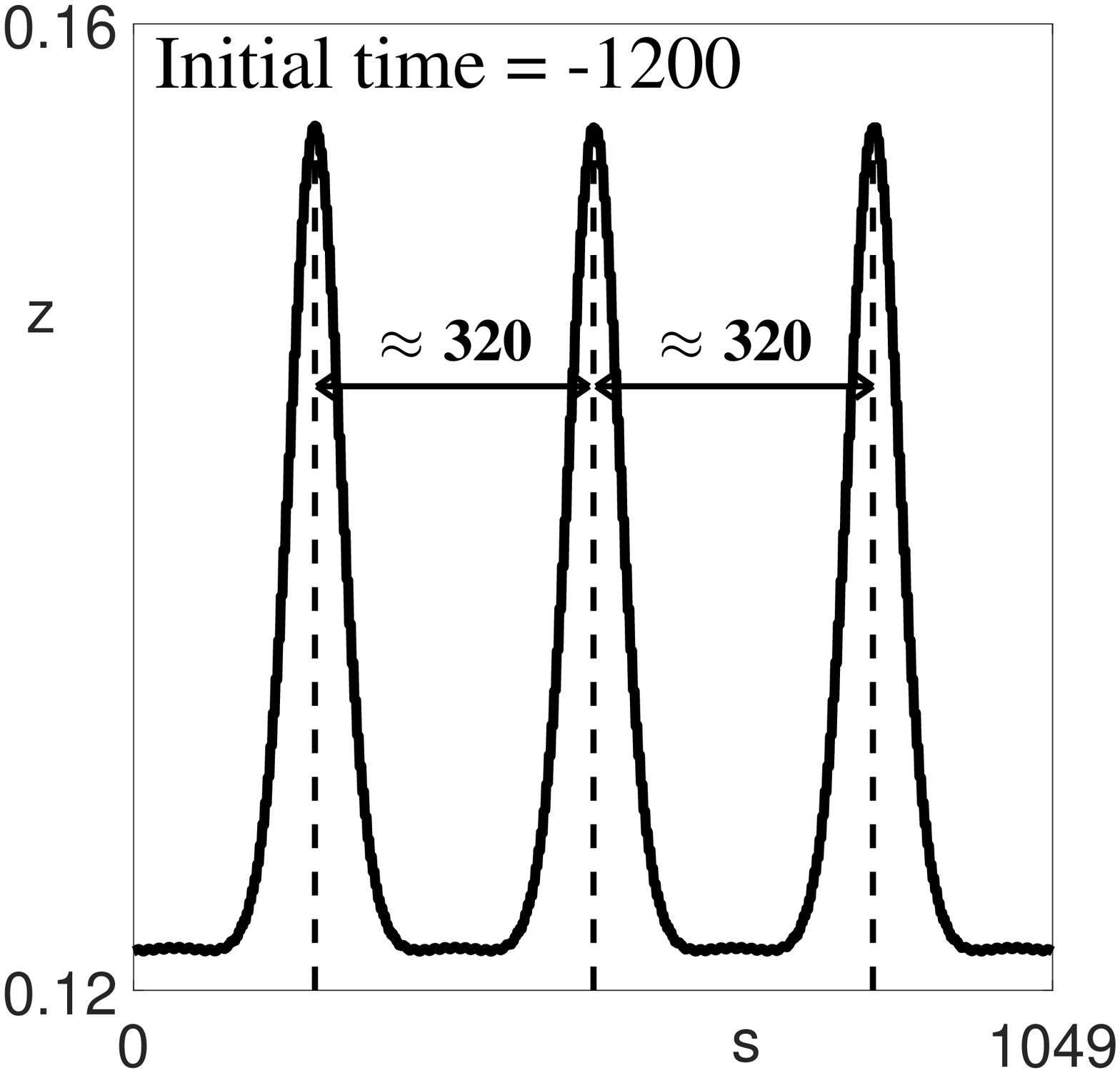}\label{fig:A-breather-init}}
    \caption{(Color online) Transmission of a gray soliton (a,b) on an s-periodic domain of $30 \times 243$ sites with initial amplitude $\Lambda=1.705i$, wavenumber $k_0=2.146$, frequency $\alpha_0=13.616$, and parameter $\psi=\pi/3$. The profile is seen initially (blue-solid) and after a long time interval (red-dashed). The appearance of an Akhmediev breather in (c,d) at $t\approx1200$, with wavenumber $k_0=2.334$, frequency $\alpha_0=12.113$, and modulation parameter $\phi=0.3$. }
\end{figure} 

Let us first consider another traveling wave solution to the 1D NLS which is the gray soliton. The gray soliton appears as localized dips in energy, however unlike the dark soliton, it is not limited to a wave minimum of strictly zero amplitude. The gray soliton envelope takes a form similar to that of the dark soliton and is given explicitly as
\begin{myequation} \label{eq:graysoliton-envelope}
C_G(S;\psi)=\Lambda\left[\cos\psi+i\sin\psi\tanh\tilde{\psi}\right],
\end{myequation}with $\tilde{\psi}=\epsilon\sin\psi(S-S_0)$ and $\psi$ an arbitrary real parameter. We see from the expression for the envelope $C_G(S;\psi)$ that the dip in energy is then described by the term $\sin\psi$ and choosing $\psi=\pi/2$ recovers the dark solution (\ref{eq:darksoliton-envelope}). Figure S\ref{fig:grey-soliton} shows the space-time evolution of a gray soliton initialized with wavenumber $k_0=2.146$, frequency $\alpha_0=13.616$, and $\psi=\pi/3$ on a periodic domain. We see that the evolution of the gray soliton is in fact similar to the dark soliton for the same wavenumber which is as expected. It is also seen from Fig. S\ref{fig:profile-grey} that the profile of the gray soliton remains undistorted throughout the evolution and even after a long time interval the background field has not decayed. The localized dip in energy has only decayed to 96\% of its original minimum amplitude which was also seen in the dark soliton case (Fig. \ref{fig:profile-dark}) as expected. 

Next let us consider another breather solution to the NLS known as the Akhmediev breather \cite{Akhmediev-etal-1987}. In contrast to the K-M soliton which is time-periodic, the Akhmediev breather is a spatially periodic solution that is localized in time. Like with the Peregrine solution, we consider initializing the solution with $t_P=1200$ so to see the growth and decay of the peaks of the breather. Due to this we use the initial condition (\ref{eq:general-initial-condition}) where the scalar envelope function is given by the Akhmediev solution \cite{Dysthe-Trulsen-1999}, i.e.
\begin{myequation}\label{eq:Abreather-envelope}
C_A(S;\phi)=\Lambda\left[ \frac{\cosh(\Omega\tilde{t}_P+2i\phi)-\cos\phi\cos\bar{S}}{\cosh(\Omega\tilde{t_P})-\cos\phi\cos\bar{S}}\right] e^{i\tilde{\theta}} ,
\end{myequation}where $\tilde{\theta}=\alpha_0 t_P-\tilde{t}_P$, $\tilde{t}_P=\epsilon^2\alpha_0''t_P$, $\bar{S}=P\epsilon(S-S_0)$, $P=2\sin\phi$, and $\Omega=\sin(2\phi)$. In fact, one can recover the K-M solution from the Akhmediev expression (\ref{eq:Abreather-envelope}) by using the parameter transformation $\phi \rightarrow i\phi$. Figure. S\ref{fig:A-breather} shows us the appearance of an Akhmediev breather where we have initialized the time-localized structure well before its maximum amplitude. It is clear to see the spatially periodic nature of the breather as all peaks form their maximum amplitude at $t\approx t_P$ and then decay away beyond this time. Figure. S\ref{fig:A-breather-init} shows that the initialized peaks are periodic approximately every 320 sites away from the position $S_0$, which is consistent with the analytical solution (\ref{eq:Abreather-envelope}).    

\end{document}